\documentclass[pdflatex,sn-nature]{sn-jnl}
\usepackage{graphicx} 
\usepackage{subcaption}
\usepackage{appendix}
\usepackage{booktabs}
\usepackage{placeins}
\usepackage{longtable}
\usepackage{MnSymbol}
\usepackage{algorithm}
\usepackage{algpseudocode}

\begin{document}

\title[Carbon Burden of Infrastructure]{Multiscale Carbon Burden of Infrastructure in the United States}

\author*[1]{\fnm{Jason} \sur{Hawkins}}\email{jfhawkin@ucalgary.ca}

\affil*[1]{\orgdiv{Department of Civil Engineering}, \orgname{University of Calgary}, \orgaddress{\street{2500 University Dr NW}, \city{Calgary}, \postcode{T2N 1N4}, \state{Alberta}, \country{Canada}}}

\abstract{Anthropogenic greenhouse gas (GHG) emissions vary spatially with development patterns, climate, economic structure, and energy systems. Using Vulcan v4.0 fossil-fuel CO2 (FFCO2) data for the United States at 1-km resolution, this study examines how land use and infrastructure shape emissions at local and metropolitan scales. I combine doubly robust Bayesian Additive Regression Tree estimators with multi-treatment spatial regression models to identify local and spillover effects by sector. A key methodological contribution is treating transportation emissions as production-based, capturing infrastructure carbon burden rather than household-attributed travel demand. Results show strong scale dependence and spatial interaction: in the preferred heteroscedasticity-robust SLX+SEM model with a 10-km distance band, residual spatial autocorrelation declines substantially. Local roadway design is a strong negative predictor of transportation FFCO2, while neighbouring land use diversity exceeds local diversity. In residential sectors, higher local density is consistently associated with lower per-capita emissions. These findings support coordinated multi-scale mitigation policy in the United States.}


\maketitle


\section{Introduction}
Engineering and economic policy focus on influencing human activity through technology, investment, and behavioural shifts to reduce GHG emissions and maximize societal well-being. However, quantifying emissions and their spatial variation at sufficiently granular scales to inform policy has historically faced data constraints. Most analyses rely on proxy measures: total vehicle kilometres travelled (VKT) may be estimated by mode, with mode-specific emissions factors applied to estimate regional transportation GHG emissions. These approaches mask important within-region heterogeneity that could inform targeted interventions.

Recent advances in data availability and processing have produced high-fidelity GHG emissions estimates at detailed spatial and temporal resolutions. These data leverage atmospheric inversion techniques to validate bottom-up sectoral estimates against top-down ensembles of atmospheric carbon-14 (C-14) measurements \cite{basu_estimating_2020}. The Vulcan dataset now provides fossil-fuel CO2 (FFCO2) emissions at 1-km grid resolution for the entire United States. However, effectively leveraging these data for climate policy analysis requires sound grounding in the urban planning and econometrics literature.

As of 2007, more than half the world's population lives in urban areas \cite{Ritchie_Samborska_Roser_2024}. In the United States, the urbanization rate exceeds 83 percent, up from 64 percent in 1950 \cite{Ritchie_Samborska_Roser_2024}. Urban form is therefore foundational to GHG reduction strategies \cite{Ramaswami_Fang_Tabory_2020}. Increasing density enables more efficient public transit and provides access to goods and services within feasible walking and cycling distances. Compact urban form also reduces land, material, and energy consumption for buildings.

While population density serves as a common proxy for built form, many other features affect VKT, mode choice, and associated emissions. Ewing and Cervero classify these features into five categories, termed ``the 5Ds of land use": population density, land use diversity (mix of uses), street design (network connectivity), distance to transit, and destination accessibility \cite{Ewing_Cervero_2010}. Compact cities scoring high on these dimensions have 20--40 percent less VKT than low-density, sprawling cities \cite{ewing_compactness_2014}.

\begin{figure}[!ht]
    \centering
    \includegraphics[width=0.5\textwidth]{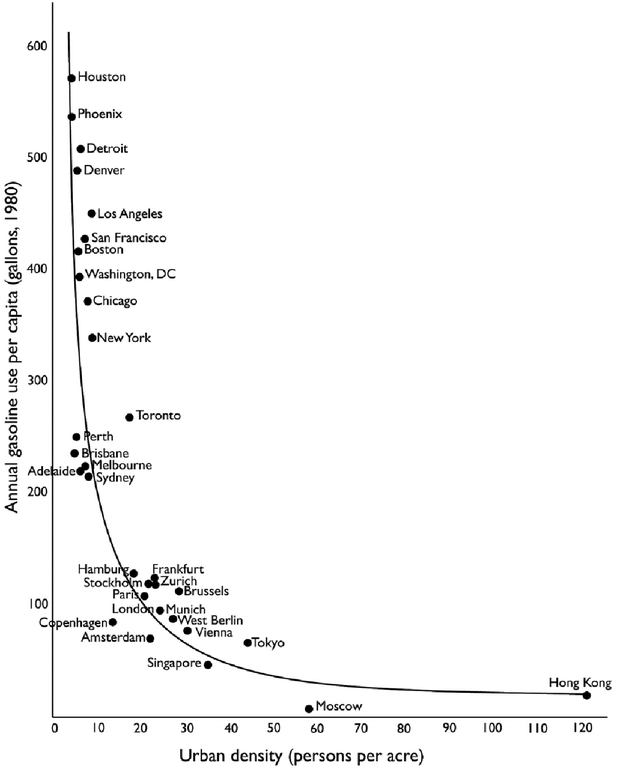}
    \caption{Annual gasoline consumption with respect to urban density \cite{Newman_Kenworthy_1989}.}
    \label{fig:kenworthy_newman}
\end{figure}

The relationships among these features are complex. Duranton and Turner estimate an elasticity of VKT with population density between -0.07 and -0.10, but their density measure does not isolate effects of other built form features \cite{Duranton_Turner_2018}. Meta-analyses find that destination accessibility has the largest impact on VKT among the 5Ds \cite{Ewing_Hamidi_2014,Stevens_2017}. Stevens reports that the elasticity of VKT with density alone is only -0.04, whereas the average effect of density as a proxy for all 5Ds may reach -0.3 \cite{Stevens_2017}. Newman and Kenworthy conducted a seminal cross-national study finding that high-density Asian cities consume less transportation energy per capita than lower-density North American cities (Figure \ref{fig:kenworthy_newman}) \cite{Newman_Kenworthy_1989}. Understanding these relationships across scales remains a central challenge.

Research in Helsinki, Finland finds that households in the urban core have higher life-cycle emissions than their suburban counterparts when considering transportation, dwellings, and goods consumption \cite{Heinonen_Junnila_2011}. Their results give 14.7 and 12.0 life-cycle tCO2eq per capita for urban core and suburban households, respectively. However, this difference likely relates to differences in the pattern of household goods consumption. Rather than being a function of built form, the difference is largely driven by urban core households having higher incomes (about 33 percent higher than suburban households). Translating these results to the US also faces significant challenges as the energy consumption profiles of American and European cities are quite different \cite{Mindali_Raveh_Salomon_2004}. Most US cities exhibit the opposite relationship between income and density, and this effect is likely magnified by suburban households consuming goods at a higher rate than urban households due to the larger size of their dwellings \cite{Murphy_2018}. With only one-fifth of dwellings being single-family detached \cite{Heinonen_Junnila_2011}, the suburbs of Helsinki have similar land use patterns to many US urban cores. Wiedenhofer et al. give a more comparable example using data from Australian cities \cite{Wiedenhofer_Lenzen_Steinberger_2013}. In that study, direct GHG emissions were higher in the suburbs due to longer travel times by private vehicle. However, indirect GHG emissions from goods consumption were generally lower for these households because they spend a larger portion of their household income on transportation costs, leaving less income available for goods consumption.

Moving beyond density to holistic compact development strategies is necessary to achieve major reductions in household GHG emissions. Lee and Lee find that doubling the population-weighted density of a US metropolitan area is associated with a 48 percent reduction in household travel emissions and 38 percent reduction in residential energy emissions \cite{Lee_Lee_2014}. Critically, population density need not double uniformly---concentrating density in specific areas can achieve similar effects. If 60 percent of new growth were directed into compact areas, the cumulative effect would equal those residents driving fuel-efficient hybrid vehicles under business-as-usual development \cite{Ewing_Bartholomew_Winkelman_Walters_Chen_2009}. Another study by Lee and Lee finds strong support for planned density, which focuses development along commercial and transit corridors \cite{Lee_Lee_2020}. Suzuki et al. demonstrate this difference through a comparison of the linear density along major transit corridors in Curitiba, Brazil, and unplanned density in Sao Paulo, Brazil \cite{Suzuki_Cervero_Iuchi_2013}.

These observations raise a fundamental question: what are the relative contributions of neighbourhood-scale design features compared to metropolitan-scale changes? Consider a mixed-use neighbourhood with pedestrian-centric streets in an otherwise automobile-dependent city. Without region-wide transit or dense employment centres, residents will likely continue driving despite local walkability. Ewing et al. find that localized neighbourhood density matters less than relative accessibility across the region \cite{Ewing_Hamidi_Tian_Proffitt_Tonin_Fregolent_2018}. Pushkar et al. similarly report that regional land use structure influences transportation emissions more than neighbourhood design in Toronto \cite{Pushkar_Hollingworth_Miller_2000}.

The literature reveals two dominant perspectives. \textbf{Macro analyses} compare metropolitan areas, emphasizing climate and regulatory heterogeneity \cite{glaeser_greenness_2010,Zheng_Wang_Glaeser_Kahn_2011}. These studies tend to discount planning measures that shift neighbourhoods along the built-form continuum, sometimes implicitly suggesting population redistribution within a country to more moderate climates. \textbf{Micro analyses} focus narrowly on neighbourhood design and building typology \cite{nichols_life-cycle_2014,nichols_urban_2015,norman_comparing_2006}, examining life-cycle energy and emissions for specific density and housing decisions. These studies, by constraining their scope, miss urban scaling effects.

A central hypothesis in the present work is that both macro and micro perspectives are necessary, but neither is sufficient, to form effective urban climate policy. Drawing on urban science \cite{West_2017}, many land use and infrastructure features follow power-law relationships with population. Cities are found to exhibit agglomeration and infrastructure scale economies \cite{Bettencourt_Lobo_Helbing_West_2007}. However, existing research provides limited insight into qualitative heterogeneity across distributions---the kind of information that translates built form features into actionable policy recommendations.

Most existing studies are correlational rather than causal in their specifications, and they consider either neighbourhood features or metropolitan effects, but rarely both simultaneously. This study addresses three key gaps:
\begin{enumerate}
\item \textbf{Causal identification:} I employ doubly robust generalised propensity score estimation to address causal identification, accounting for both treatment assignment and outcome models.
\item \textbf{Multi-scale relationships:} I include land use variables defined at both local (census block group) and metropolitan (core-based statistical area) scales. Models are developed to examine how effects operate across spatial hierarchies within a consistent model framework.
\item \textbf{Hosted transportation emissions:} I provide production-based CO2 emissions allocation for the transportation sector as a complement to consumption-based allocations in the literature.
\item \textbf{Sectoral disaggregation:} I distinguish among transportation, residential electricity, and residential non-electricity energy sources to understand differential effects of land use across emissions sources.
\end{enumerate}

Importantly, I examine production-based transportation emissions as allocated in Vulcan v4.0, which assigns emissions to their point of generation rather than household origins. This framing shifts analysis towards understanding the \textbf{carbon burden of infrastructure} rather than consumption-based allocation. As discussed in Section \ref{subsec:trans_res}, this reveals spatial mismatches whereby urban centres ``host" emissions for surrounding regions. The empirical estimates are grounded in US-specific emissions accounting, urban form, and transit availability; therefore, the results should not be interpreted as directly generalisable to other regions. Instead, the study offers a transferable empirical approach for regions where comparable gridded emissions, land use, and infrastructure data exist. Several data sources provide global GHG estimates by sector at 0.1 degree resolution \cite{Crippa_etal_2024,Dou_al_2023}, while many researchers and cities have compiled disaggregate GHG estimates for individual cities \cite{Welegedara_Agrawal_2024}.

\section{Results}\label{sec:model_res}
I focus my analysis on three sources of fossil fuel-based CO2 (FFCO2) emissions: total transportation, residential electricity, and residential non-electricity energy. The data source (Vulcan v4.0, see ``Data" for more details) provides annual FFCO2 estimates for 1-km gridcells, which I convert to census block groups (CBG) to align with land use and urban form data available from the EPA. Vulcan v4.0 uses a Scope 1 allocation approach in which emissions are assigned to their point of production. A recent extension provides Scope 2 consumption-based electricity emissions allocation to the points of demand \cite{Gurney_Dass_Kato_Mitra_Nematchoua_2024}. I present two sets of model estimates for each emissions source. First, doubly robust (DR) average treatment effect (ATE) posterior distributions are constructed for each of the relevant 5D land use variables measured by CBG. A DR approach accounts for both the probability of land use treatment assignment and the outcome model, ensuring unbiased estimates even if one model is misspecified. Second, log-linear regression models are estimated that incorporate all of the relevant 5D land use variables for CBG and core-based statistical areas (CBSA) spatial units. I assume that CBG represent neighbourhoods, while CBSA represent metropolitan areas. The 5D land use variables were developed for transportation vehicle kilometres travelled (VKT) analysis, so I consider all five to influence transportation FFCO2 emissions. Residential FFCO2 emissions are assumed to be influenced by population density and land use diversity, which provide surrogate measures for dwelling area, shared walls, and other relevant factors. I assume that roadway design, distance to transit, and destination accessibility have no influence on residential energy use rates beyond their correlations with density and diversity variables.

\subsection{Transportation Treatment Effects}\label{subsec:trans_res}
Before presenting results, it is critical to understand the allocation methodology used in Vulcan v4.0. Unlike most transportation-emissions studies that allocate emissions to household locations (consumption-based), Vulcan assigns transportation emissions to their point of production-where vehicles physically travel and emit FFCO2. This distinction fundamentally affects interpretation.

In practical terms, a dense urban neighbourhood hosting a major arterial road will show high transportation emissions even if its residents drive infrequently. The emissions reflect through-traffic, commuters from suburbs, and commercial vehicles serving the broader region. Conversely, a low-density suburban neighbourhood may show low production-based emissions despite residents generating substantial vehicle travel that occurs elsewhere.

I term these ``hosted emissions" to distinguish them from household-generated emissions. This framing reorients analysis towards the \textbf{carbon burden infrastructure places on communities} rather than household travel behaviour. The production-based perspective reveals important environmental justice dimensions: urban core residents experience air quality impacts and ``host" emissions on behalf of suburban residents who may be the actual generators of travel demand.

This allocation bias should be minimal at the CBSA (metropolitan) scale, since most personal travel occurs within the metropolitan region. I partially control for allocation bias at the CBG (neighbourhood) scale using auxiliary variables from the EPA Smart Location Database, including worker-based emissions and commute vehicle miles travelled. Section \ref{sec:discussion} discusses more robust approaches to address this limitation in future work.
With this framework in mind, the following results should be interpreted as the relationship between land use features and the spatial distribution of transportation infrastructure carbon burden, not necessarily household travel emissions. This analysis provides a complementary perspective to consumption-based analyses in the literature. Valid inference for transportation emissions concerns infrastructure carbon burden and air quality exposure rather than household behaviour, with applications to congestion pricing (CP), low-emission zones (LEZ), and facility siting decisions discussed further in Section 3.2.

\begin{figure}[htbp]
    \centering
    \includegraphics[width=\textwidth]{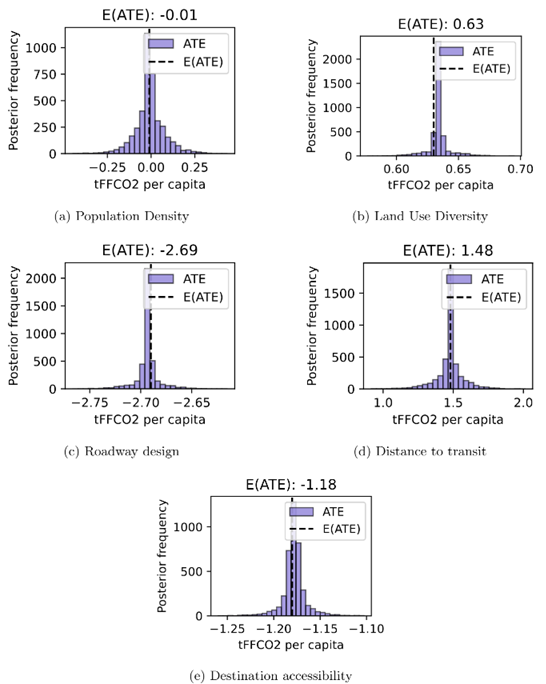}
    \caption{Doubly robust average treatment effects (ATE) of land use variables on transportation FFCO2 emissions for a change from the 25$^{th}$ to 75$^{th}$ percentile treatment value. Panels show effects for population density, land use diversity, roadway design, distance to transit, and destination accessibility.}
    \label{fig:tran_grid}
\end{figure}

The single-treatment DR average treatment effect (ATE) transportation results are provided in Figure \ref{fig:tran_grid} as changes in tonnes of emissions (tFFCO2) for a change in the treatment variable from its 25$^{th}$ to 75$^{th}$ percentile value. By taking a Bayesian approach to inference in the DR models, I specify distributions on the ATEs rather than point estimates. Thus, results are robust to uncertainty in the underlying data. These results are interpretable as the \emph{total effect} of each land use feature after balancing covariates but not conditioning on other features within the ``5Ds" framework. That is, the ATEs represent total treatment effects where other land use variables are treated as correlated constructs rather than confounders. I calculate the posterior of residual spatial autocorrelation using Moran's I statistic as 0.28, with negligible variation across posteriors or treatments.

Local (CBG) population density shows minimal effect on production-based transportation emissions in the DR model, with a posterior distribution centred near zero. This contrasts with the joint model results (Table \ref{tab:tran_joint_effects}), discussed below. Land use diversity (measured by employment entropy) tends to substantially increase hosted emissions. Roadway design (measured as facility miles per square mile) demonstrates a strong negative effect of -2.69 tFFCO2 per capita. This suggests that denser street networks - characteristic of walkable neighbourhoods with shorter block lengths and more intersections - tend to have lower emissions due to lower traffic volumes. Intuitively, being further from transit increases production-based tFFCO2 in the DR model. Higher destination accessibility shows a negative effect on hosted emissions, suggesting that access to jobs is not necessarily associated with proximity to major roadways.

Taken together, these production-based results reveal a spatial mismatch in emission distributions: neighbourhoods with desirable urban form features (density and mixed uses) tend to host the transportation infrastructure that generates regional mobility, thereby bearing higher localized emissions even though their residents may generate less vehicle travel.

My findings confirm consumption-based results by Zahabi, who report a 5.8\% reduction in overall greenhouse gas (GHG) emissions for a 10\% increase in transit accessibility \cite{Zahabi_Miranda-Moreno_Patterson_Barla_Harding_2012}. More broadly, studies of compact urban form consistently show lower GHG emissions when measured from a demand perspective \cite{Anderson_Kanaroglou_Miller_1996,Christen_Coops_Crawford_Kellett_Liss_Olchovski_Tooke_2011,Dhakal_2009,Kennedy_Steinberger_Gasson_Hansen_Hillman_Havránek_Pataki_Phdungsilp_Ramaswami_Mendez_2009}. My findings using national data capture the reality that 88 percent of US metropolitan areas lack reportable transit service (per EPA-SLD). In these regions, density and mixed uses may concentrate traffic without providing viable low-carbon alternatives, creating the ``hosting" patterns I document.

In addition to the independent DR estimators shown above, I also estimate a joint model for all outcomes, using a simpler frequentist linear regression to avoid issues encountered in attempting to estimate a more flexible Bayesian model. The model includes the treatment estimators at the CBG and CBSA spatial scales, as well as the propensity score (PS) and interactions with treatments at the CBG scale. I use z-score standardisation to reduce the condition number (i.e., control numerical stability). The Durbin-Watson statistic indicates non-normal residual errors in this model. The adjusted-$R^2$ suggests moderate prediction accuracy. Standard errors are clustered at the state level to partially address robustness of results given the Durbin-Watson statistic. A second specification includes spatial eigenvectors to control for residual spatial autocorrelation. In the joint model, 5D effects are treated as \emph{partial effects} holding other 5D variables fixed. Residual Moran's I statistics indicate substantial spatial autocorrelation in the baseline specifications, consistent with omitted spatial processes. I therefore estimate SLX+SEM models under both Queen contiguity and inverse-distance (10 km) weights. The Queen model appears misspecified for transportation activity: residual Moran's I remains high (0.384, slightly worse than 0.368 in the spatial eigenvector model). In contrast, the 10 km specification reduces residual Moran's I to 0.174, implying that transport emissions operate within a broader commuting radius rather than only along shared borders. Heteroscedasticity-robust inference supports this scale choice: roadway design and land use diversity remain highly significant even under conservative variance correction. The spatial-error coefficient is also high in the preferred 10 km model ($\lambda=0.712$), with alternative HET variants around 0.60, which is consistent with successful filtering of strong regional unobservables from the land use coefficients.

The joint model is set up as a log-linear model with standardised covariates, so parameters are interpretable as percent changes in FFCO2 for a 1 SD change in the treatment variable. I use the SLX+SEM (10 km) model as representing the most robust results. Increasing CBSA population density by 1 SD is associated with an 11.3 percent increase in FFCO2. All CBSA effects are statistically insignificant after accounting for local treatments and propensity scores. Population density by CBG has the opposite sign as for the CBSA but a larger negative effect of 19.5 percent for a 1 SD increase in population density. The sign is consistent with the doubly robust estimate. The only variable that exhibits a change in sign relative to the DR results is destination accessibility.

Comparing joint model results with those from the DR models, we can assess overlapping mechanisms among the 5D variables. Roadway design maintains a strong and negative effect in the joint model. Destination accessibility has inconsistent effects between the single treatment and joint models. When accessibility is measured conditional on other features (joint model), the effect is strongly positive, suggesting that more accessible CBGs tend to have higher FFCO2 emissions. This treatment effect is partially balanced by negative spatial spillovers. The Queen contiguity model shows the opposite results, with a weak direct effect and stronger positive spatial spillover. Stevens argues that density serves as a proxy for multiple correlated urban form features, which could explain some of the difference in results between the single and joint treatment models \cite{Stevens_2017}.

\begin{table}[htbp]

\centering
\caption{Joint treatment effects of land use variables on transportation FFCO$_2$ emissions}
\label{tab:tran_joint_effects}
\small
\begin{tabular}{@{}lcccc@{}}
\toprule
& \textbf{Base OLS}
& \textbf{Spatial EV}
& \textbf{\shortstack{SEM+\\SLX (Q)}}
& \textbf{\shortstack{SEM+\\SLX (10 km)}} \\
\textbf{Variable}
& \textbf{Coef. (p)}
& \textbf{Coef. (p)}
& \textbf{Coef. (p)}
& \textbf{Coef. (p)} \\
\midrule
Constant & 1.407 (0.00) & 1.408 (0.00) & 1.334 (0.00) & 1.579 (0.00) \\
\textit{Regional Context (CBSA)} & & & & \\
Total population (CBSA) & -0.054 (0.30) & -0.054 (0.29) & -0.030 (0.62) & -0.044 (0.50) \\
Average population density (CBSA) & 0.017 (0.96) & 0.021 (0.95) & 0.217 (0.58) & 0.107 (0.80) \\
Average land use diversity (CBSA) & 0.010 (0.25) & 0.009 (0.26) & 0.005 (0.66) & 0.015 (0.18) \\
Average roadway design (CBSA) & 0.009 (0.50) & 0.009 (0.51) & -0.012 (0.46) & -0.018 (0.29) \\
Average distance to transit (CBSA) & -0.001 (0.84) & -0.001 (0.86) & 0.003 (0.75) & 0.003 (0.76) \\
Average destination accessibility (CBSA) & 0.047 (0.32) & 0.047 (0.32) & 0.013 (0.81) & 0.071 (0.22) \\
\midrule
\textit{Propensity Scores (PS)} & & & & \\
Population density PS & -0.102 (0.02) & -0.102 (0.02) & -0.106 (0.00) & -0.114 (0.00) \\
Land use diversity PS & 0.021 (0.04) & 0.021 (0.04) & -0.003 (0.43) & 0.007 (0.05) \\
Roadway design PS & -0.003 (0.78) & -0.003 (0.78) & 0.006 (0.07) & 0.001 (0.82) \\
Distance to transit PS & 0.029 (0.14) & 0.029 (0.14) & 0.025 (0.00) & 0.026 (0.00) \\
Destination accessibility PS & 0.022 (0.19) & 0.022 (0.20) & 0.028 (0.00) & 0.036 (0.00) \\
\midrule
\textit{Local Treatments (CBG)} & & & & \\
Population density (CBG) & -0.187 (0.00) & -0.187 (0.00) & -0.197 (0.00) & -0.194 (0.00) \\
Land use diversity (CBG) & 0.123 (0.00) & 0.123 (0.00) & 0.105 (0.00) & 0.126 (0.00) \\
Roadway design (CBG) & -0.449 (0.00) & -0.449 (0.00) & -0.458 (0.00) & -0.475 (0.00) \\
Distance to transit (CBG) & -0.001 (0.98) & -0.001 (0.98) & 0.039 (0.00) & 0.067 (0.00) \\
Destination accessibility (CBG) & 0.324 (0.01) & 0.324 (0.01) & 0.095 (0.00) & 0.447 (0.00) \\
\midrule
\textit{Treatment $\times$ PS Interactions} & & & & \\
Population density $\times$ PS (CBG) & 0.029 (0.15) & 0.029 (0.15) & 0.040 (0.02) & 0.034 (0.03) \\
Land use diversity $\times$ PS (CBG) & 0.039 (0.00) & 0.039 (0.00) & 0.019 (0.00) & 0.028 (0.00) \\
Roadway design $\times$ PS (CBG) & 0.007 (0.32) & 0.006 (0.32) & 0.009 (0.00) & 0.006 (0.01) \\
Distance to transit $\times$ PS (CBG) & 0.005 (0.77) & 0.005 (0.77) & 0.005 (0.04) & 0.000 (0.96) \\
Destination accessibility $\times$ PS (CBG) & 0.015 (0.40) & 0.015 (0.40) & 0.003 (0.33) & 0.011 (0.00) \\
\midrule
\textit{Spatial Eigenvectors} & & & & \\
Spatial eigenvector 1 & --- & 0.030 (0.75) & --- & --- \\
Spatial eigenvector 2 & --- & 0.061 (0.48) & --- & --- \\
Spatial eigenvector 3 & --- & 0.135 (0.16) & --- & --- \\
Spatial eigenvector 4 & --- & -0.066 (0.50) & --- & --- \\
Spatial eigenvector 5 & --- & 0.082 (0.46) & --- & --- \\
\midrule
\textit{Spatial Lags (SLX)} & & & & \\
$W$ Population density & --- & --- & -0.023 (0.30) & 0.038 (0.44) \\
$W$ Land use diversity & --- & --- & 0.181 (0.00) & 0.223 (0.00) \\
$W$ Roadway design & --- & --- & 0.031 (0.00) & 0.123 (0.00) \\
$W$ Distance to transit & --- & --- & -0.086 (0.00) & -0.144 (0.00) \\
$W$ Destination accessibility & --- & --- & 0.253 (0.00) & -0.190 (0.00) \\
\midrule
\textit{Diagnostics} & & & & \\
Lambda ($\lambda$) & --- & --- & 0.681 (0.00) & 0.706 (0.00) \\
Residual Moran's I & 0.370 & 0.368 & 0.384 & 0.174 \\
Adj-$R^2$ & 0.268 & 0.268 & 0.266 & 0.259 \\
Condition number & 4180 & 4180 & --- & --- \\
Durbin-Watson statistic & 1.35 & 1.35 & --- & --- \\
\bottomrule
\end{tabular}
\footnotetext{\textit{Note:} OLS uses state-clustered SE. SEM models use GMM Heteroscedastic-robust SE. "Q" denotes Queen contiguity weights.}

\end{table}

\FloatBarrier

\subsection{Residential Electricity Treatment Effects}
\begin{figure}[htbp]
    \centering
        \includegraphics[width=\textwidth]{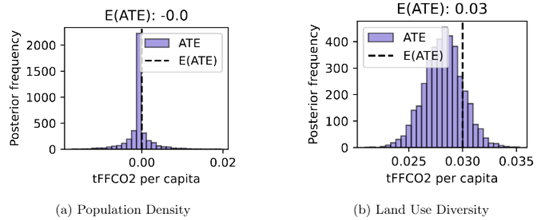}
        \caption{Doubly robust average treatment effects (ATE) of land use variables on residential electricity FFCO2 emissions for a change from the 25$^{th}$ to 75$^{th}$ percentile treatment value. Panels show effects for population density and land use diversity.}

    \label{fig:elec_grid}
\end{figure}

Because Vulcan v4.0 allocates residential emissions to stationary point sources, I expect fewer allocation biases than in transportation. I also assume that only population density and land use diversity affect these outcomes. I find that population density tends to have a negligible effect on residential electricity emissions (Figure \ref{fig:elec_grid}a). The diversity effect is larger (0.03 tFFCO2 per capita) using a 5-digit NAICS code employment entropy measure (Figure \ref{fig:elec_grid}b). I calculate the posterior of residual spatial autocorrelation using Moran's I statistic as 0.49, with negligible variation across posteriors or treatments. Unlike transportation emissions, residential electricity emissions do not have a theoretical spatial spillover mechanism in the treatment effect because emissions are assigned to fixed dwellings rather than mobile network activity; the remaining spatial structure is therefore interpreted as spatially correlated residual variation, discussed further in Section \ref{sec:discussion}.

Table \ref{tab:elec_joint_effects} shows results for the joint model. For both the residential models, I do not expect spatial spillovers as emissions are produced at fixed locations. However, I include model specifications for each with spatial eigenvectors (SEV) as a basic robustness against unobserved residual spatial errors. Residual Moran's I remains 0.258 in the residential electricity model, indicating remaining spatial autocorrelation in the errors. I interpret this as evidence of spatially correlated omitted factors, such as climate, housing vintage, and income, rather than interference among neighbouring CBG treatment effects. The model suggests minimal effect of CBSA population density or land use diversity on FFCO2 emissions, but statistically significant local effects of -6.8\% and 3.7\% for CBG population density and land use diversity, respectively. These results closely align with Lee and Lee \cite{Lee_Lee_2014}. The consistent signs between single and joint models indicate that results are robust to specification. Overall, more diverse CBGs tend to have higher residential electricity emissions, while density tends to decrease emissions.

\begin{table}[htbp]

\centering
\caption{Joint treatment effects of land use variables on residential electricity FFCO$_2$ emissions}
\label{tab:elec_joint_effects}
\small
\begin{tabular}{@{}lcc@{}}
\toprule
& \textbf{Base Model} & \textbf{\shortstack{With Spatial\\Eigenvectors}} \\
\textbf{Variable} & \textbf{Coef. (p)} & \textbf{Coef. (p)} \\
\midrule
Constant & -0.240 (0.00) & -0.240 (0.00) \\
\textit{Regional Context (CBSA)} & & \\
Total population (CBSA) & 0.006 (0.79) & 0.005 (0.80) \\
Average population density (CBSA) & -0.138 (0.53) & -0.135 (0.54) \\
Average land use diversity (CBSA) & -0.015 (0.23) & -0.015 (0.23) \\
\midrule
\textit{Propensity Scores (PS)} & & \\
Population density PS & 0.013 (0.24) & 0.013 (0.24) \\
Land use diversity PS & -0.024 (0.00) & -0.024 (0.00) \\
\midrule
\textit{Local Treatments (CBG)} & & \\
Population density (CBG) & -0.068 (0.01) & -0.068 (0.01) \\
Land use diversity (CBG) & 0.037 (0.00) & 0.037 (0.00) \\
\midrule
\textit{Treatment $\times$ PS Interactions} & & \\
Population density × PS (CBG) & 0.001 (0.87) & 0.001 (0.87) \\
Land use diversity × PS (CBG) & -0.035 (0.00) & -0.035 (0.00) \\
\midrule
\textit{Spatial Eigenvectors} & & \\
Eigenvector 0 & --- & -0.048 (0.59) \\
Eigenvector 1 & --- & -0.007 (0.95) \\
Eigenvector 2 & --- & 0.143 (0.10) \\
Eigenvector 3 & --- & -0.139 (0.03) \\
Eigenvector 4 & --- & 0.006 (0.94) \\
\midrule
\textbf{Diagnostics} & & \\
Residual Moran's I & 0.258 (0.00) & 0.258 (0.00) \\
Adj-$R^2$ & 0.648 & 0.648 \\
Condition number & 2610 & 2610 \\
Durbin–Watson statistic & 1.16 & 1.16 \\
\bottomrule
\end{tabular}
\footnotetext{\textit{Note:} Standard errors are clustered at the state level. CBSA fixed effects omitted for clarity. High condition numbers are due to CBSA FE.}

\end{table}

\FloatBarrier

\subsection{Residential Energy Treatment Effects}
\begin{figure}[htbp]
    \centering
    \includegraphics[width=\textwidth]{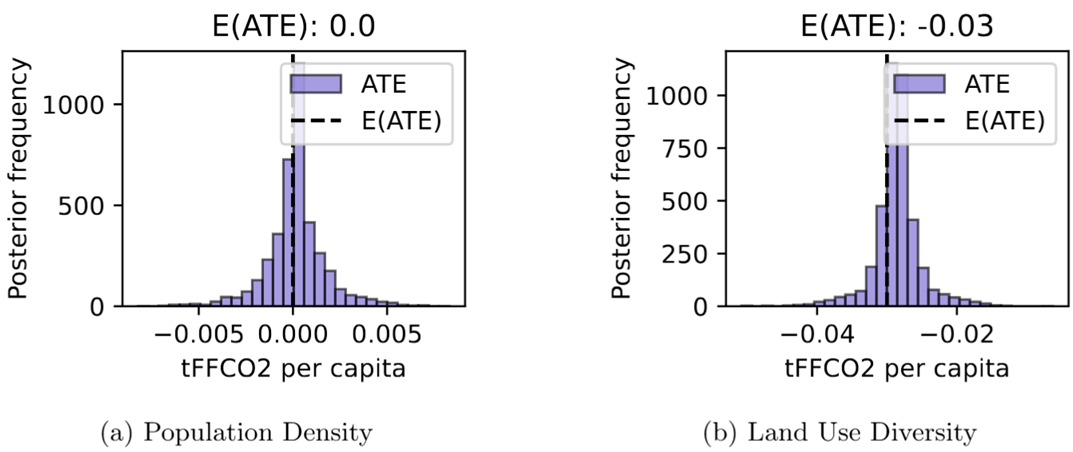}
        \caption{Doubly robust average treatment effects (ATE) of land use variables on residential energy FFCO2 emissions for a change from the 25$^{th}$ to 75$^{th}$ percentile treatment value. Panels show effects for population density and land use diversity.}
    \label{fig:res_grid}
\end{figure}

For the same reasons given above, residential energy results should not be subject to allocation bias. The single-treatment DR results indicate that density and land use diversity have no meaningful effect on tFFCO2 emissions (Figure \ref{fig:res_grid}). Residual spatial autocorrelation remains in the joint residential non-electricity energy model (Moran's I = 0.431), but this is again interpreted as spatially correlated residual heterogeneity rather than a structural spillover process in household heating fuel consumption.

Multi-treatment results in Table \ref{tab:res_joint_effects} show minimal effect from CBSA population density and land use diversity on FFCO2 emissions for residential energy. However, local (CBG) population density has a strong negative (-12.0\% per 1 SD) effect on these emissions. This aligns with previous findings that denser neighbourhoods reduce per capita residential energy use \cite{Dhakal_2009,Kennedy_Steinberger_Gasson_Hansen_Hillman_Havránek_Pataki_Phdungsilp_Ramaswami_Mendez_2009}. As in the electricity model, the similarity between the single-treatment DR estimates and the joint regression results supports the robustness of the treatment-effect interpretation despite residual spatial autocorrelation.

\begin{table}[htbp]

\centering
\caption{Joint treatment effects of land use variables on residential energy FFCO$_2$ emissions}
\label{tab:res_joint_effects}
\small
\begin{tabular}{@{}lcc@{}}
\toprule
& \textbf{Base Model} & \textbf{\shortstack{With Spatial\\Eigenvectors}} \\
\textbf{Variable} & \textbf{Coef. (p)} & \textbf{Coef. (p)} \\
\midrule
Constant & -1.076 (0.00) & -1.077 (0.00) \\
\textit{Regional Context (CBSA)} & & \\
Total population (CBSA) & 0.029 (0.17) & 0.029 (0.17) \\
Avg. population density (CBSA) & -0.033 (0.89) & -0.033 (0.89) \\
Avg. land use diversity (CBSA) & -0.006 (0.65) & -0.007 (0.63) \\
\midrule
\textit{Propensity Scores (PS)} & & \\
Population density PS & 0.015 (0.62) & 0.015 (0.62) \\
Land use diversity PS & 0.014 (0.07) & 0.014 (0.06) \\
\midrule
\textit{Local Treatments (CBG)} & & \\
Population density (CBG) & -0.120 (0.01) & -0.120 (0.01) \\
Land use diversity (CBG) & 0.008 (0.18) & 0.008 (0.18) \\
\midrule
\textit{Treatment $\times$ PS Interactions} & & \\
Population density × PS (CBG) & -0.029 (0.15) & -0.029 (0.15) \\
Land use diversity × PS (CBG) & 0.010 (0.05) & 0.010 (0.05) \\
\midrule
\textit{Spatial Eigenvectors} & & \\
Eigenvector 0 & --- & 0.047 (0.59) \\
Eigenvector 1 & --- & -0.175 (0.03) \\
Eigenvector 2 & --- & 0.038 (0.67) \\
Eigenvector 3 & --- & 0.089 (0.23) \\
Eigenvector 4 & --- & 0.152 (0.12) \\
\midrule
\textit{Diagnostics} & & \\
Residual Moran's I & 0.431 (0.00) & 0.431 (0.00) \\
Adj-$R^2$ & 0.638 & 0.638 \\
Condition number & 2610 & 2610 \\
Durbin–Watson statistic & 1.54 & 1.54 \\
\bottomrule
\end{tabular}
\footnotetext{\textit{Note:} Standard errors are clustered at the state level. CBSA fixed effects omitted. Condition numbers inflated by CBSA FE.}

\end{table}

\FloatBarrier

\section{Discussion}\label{sec:discussion}
This study reveals complex, scale-dependent relationships between urban form and fossil-fuel CO2 (FFCO2) emissions. Three overarching patterns emerge from the analysis:

\textbf{First, production-based and consumption-based emission allocations tell fundamentally different stories.} Using production-based transportation emissions from Vulcan v4.0, I find that neighbourhood-scale effects are heterogeneous: higher local population density is associated with lower transportation FFCO2, while greater land use diversity and destination accessibility are associated with higher hosted emissions in the preferred joint spatial specification. This contrasts with the consumption-based literature, which generally shows that residents of compact, accessible neighbourhoods generate less vehicle travel \cite{Zahabi_Miranda-Moreno_Patterson_Barla_Harding_2012,Ewing_Hamidi_Tian_Proffitt_Tonin_Fregolent_2018}. Taken together, these results reinforce an environmental justice concern: urban core communities can bear disproportionate air quality and carbon burdens from infrastructure serving regional mobility, even when local travel demand may be lower.

\textbf{Second, residential emissions exhibit clear but sector-specific relationships with urban form.} At the neighbourhood (CBG) scale, higher population density is consistently associated with lower per-capita residential emissions in both sectors (about -6.8\% for electricity and -12.0\% for non-electricity energy in the joint models), aligning with prior research \cite{Dhakal_2009,Kennedy_Steinberger_Gasson_Hansen_Hillman_Havránek_Pataki_Phdungsilp_Ramaswami_Mendez_2009}. The likely mechanism is straightforward: denser housing implies smaller units, more shared walls, and lower energy demand per resident. By contrast, metropolitan-scale (CBSA) density effects are weak and statistically insignificant in both residential models. Land use diversity shows a modest positive association for residential electricity (about +3.7\%), but limited influence for residential non-electricity energy.

\textbf{Third, roadway network density remains a dominant local predictor of lower transportation emissions, but with meaningful spatial spillovers.} In the preferred SLX+SEM (10 km) specification, higher local facility miles per square mile is associated with a large negative direct effect on production-based transportation FFCO$_2$, while the spatial lag term is positive, indicating that neighbouring network structure also shapes hosted emissions. This pattern suggests that connected street networks can reduce local concentration of traffic emissions, but policy impacts depend on coordinated network design across adjacent areas rather than isolated neighbourhood interventions.

The divergent effects across spatial scales have important implications for climate policy. Regional development patterns and transportation systems shape the baseline context within which local interventions operate. In auto-oriented metropolitan regions, even dense, walkable neighbourhoods remain embedded in systems generating high regional emissions. This dampens or counteracts the emissions-reducing benefits of local compactness.

For transportation policy, the production-based perspective reveals spatial mismatches requiring explicit management. The HET-robust 10 km results also imply a multiplier at the cluster scale: the neighbouring diversity effect (0.209) exceeds the local diversity effect (0.124), so policy should target diverse 10 km commuting-shed clusters rather than isolated block-level interventions. Policy options include:
\begin{itemize}
    \item \textbf{Congestion pricing} in urban cores and along high traffic corridors to reflect the external costs imposed on residents who host regional traffic, as in New York City \cite{Cook_Kreidieh_Vasserman_Allcott_Arora_vanSambeek_Tomkins_Turkel_2025}.
    \item \textbf{Low-emission zones} that restrict polluting vehicles in dense neighbourhoods bearing disproportionate air quality burdens, as in many European urban cores \cite{Tam_2023}.
    \item \textbf{Revenue redistribution} from transportation infrastructure user fees to communities hosting emissions-intensive infrastructure.
    \item \textbf{Mobility hub development} that concentrates park-and-ride, transit, and micromobility at regional peripheries rather than pushing all traffic through urban cores.
\end{itemize}

For residential emissions, policies should recognize that:
\begin{itemize}
\item \textbf{Neighbourhood-scale density} consistently reduces per-capita emissions in both residential sectors, supporting compact development with efficient multifamily building forms and shared infrastructure.
\item \textbf{Metropolitan-scale density} has weak and statistically insignificant effects in the joint models, so regional density targets alone are unlikely to deliver large residential emissions reductions without local design and retrofit measures.
\item \textbf{Sector-specific interventions are needed:} land use diversity is associated with modestly higher residential electricity emissions, while effects for residential non-electricity energy are limited; policy should pair compact growth with building performance standards, electrification, and weatherization in existing housing stock.
\end{itemize}

Overall, monitoring of emissions is important to track policy effectiveness and changing dynamics \cite{Assimacopoulos_etal_2025,Martín_Ortega_etal_2024}. My analysis demonstrates that monitoring must be spatially disaggregated to facilitate measurement and attribution of emissions to their sources. Policy, particularly for residential emissions, will be strongly dependent on heating and cooling loads as a function of climatic conditions \cite{Welegedara_Agrawal_2024}.

Several limitations affect interpretation and suggest future research directions:

\textbf{Production-based transportation allocation} is the most significant limitation. Vulcan v4.0 assigns emissions to their spatial point of generation. While I partially control for this using worker-based emissions and commute VMT variables, residual bias likely remains at the neighbourhood scale. Two approaches could address this:
\begin{enumerate}
    \item \textbf{Location-based service (LBS) data:} Cellular phone movement data could link trips to household origins. Trip routing algorithms could then reassign gridcell emissions to origin census block groups (CBGs). This approach is computationally intensive and expensive but would provide high-quality consumption-based estimates.
    \item \textbf{Census LODES data:} Origin-destination commuting flows could allocate commute-related emissions to household locations at lower computational cost but with less comprehensive trip coverage.
\end{enumerate}

Temporal constraints affect generalizability. This analysis uses 2019 emissions data. Post-pandemic changes in work arrangements, e-commerce, and travel behaviour may have altered some relationships. Panel analysis across multiple years could identify temporal stability of effects. Causal identification challenges also remain despite the doubly robust approach. Residential self-selection into neighbourhoods based on preferences and income creates potential bias. Instrumental variable approaches or natural experiments (policy changes, new transit lines) could strengthen causal claims. Data availability may be restricted for other regions, though 0.1 degree gridded estimates are available through the EDGAR dataset \cite{Crippa_etal_2024} and many cities produce their own GHG estimates.

The nonparametric BART models for doubly robust estimation and linear models for joint effects represent different tradeoffs between flexibility and interpretability. More sophisticated approaches (e.g., copula models capturing outcome correlations) may reveal additional insights but face steep estimation challenges with continuous multi-treatment frameworks. Scope 2 electricity emissions introduce additional uncertainty through the multiregional input--output (MRIO) modelling used to allocate grid-mix emissions to consumption locations. Validation suggests reasonable accuracy, but region-specific analyses should consider local utility generation mixes.

\section{Methods}\label{sec:methods}
\subsection{Data}\label{sec:data}
A core dataset for this work is the Vulcan v4.0 fossil fuel-based CO2 (FFCO2) estimates constructed by Gurney et al. \cite{Gurney_Liang_Patarasuk_Song_Huang_Roest_2020}. The Vulcan dataset comprises 1 km gridcell FFCO2 estimates for the US between 2010--2021. The data are further disaggregated into the following source categories: on-road, non-road, airport, commercial marine vessels, railroads, electricity production, residential non-point buildings, commercial non-point buildings, commercial point sources, industrial point sources, industrial non-point buildings, and cement. Estimates are provided at an hourly resolution, though I aggregate these estimates to annual totals. The Vulcan dataset has been refined and validated across multiple studies \cite{Gurney_Dass_Kato_Mitra_Nematchoua_2024,Gurney_Liang_Patarasuk_Song_Huang_Roest_2020,Patarasuk_Gurney_O’Keeffe_Song_Huang_Rao_Buchert_Lin_Mendoza_Ehleringer_2016}. It includes confidence bands derived from the uncertainty of the underlying input datasets. Validation includes comparison against three other estimates for the US domain: the CDIAC, EDGAR (v4.2 and v4.3) and the USEPA \cite{basu_estimating_2020}. Vulcan v4.0 and the atmospheric-derived estimate for 2010 agreed to within 3.3\% but departed significantly from the aggregated totals from other estimates \cite{Gurney_Dass_Kato_Gawuc_Aslam_Sun_2025}. Though the uncertainty bounds overlapped, the USEPA central estimate was 6\% lower than the total Vulcan v3.0 FFCO2 emissions (assume similar variations for Vulcan V4.0) \cite{Gurney_Liang_Patarasuk_Song_Huang_Roest_2020}. The ODIAC dataset provides gridded estimates comparable to Vulcan v4.0 \cite{Oda_Maksyutov_Andres_2018}. In validation work, Gurney et al. find that 2011 estimates for ODIAC and Vulcan V3.0 are 1,453.5 TgC/year and 1,553.8 TgC/year, respectively, or a difference of 7.6\% \cite{Gurney_Liang_Patarasuk_Song_Huang_Roest_2020}. The Vulcan estimates are further calibrated against atmospheric FFCO2 measurement based on NOAA's air sampling network. Gurney et al. compare Vulcan estimates to reports by 48 US cities and find these cities under-report their Scope 1 emissions by about 18.3 percent (range of -145.5 to +63.5 percent) \cite{Gurney_Liang_Roest_Song_Mueller_Lauvaux_2021}. If extrapolated to all US cities, this amounts to 123.5 percent of all California Scope 1 emissions by their estimation.

Recently, a supplemental Vulcan dataset was released for Scope 2 emissions for the 2019--2021 time period \cite{Gurney_Dass_Kato_Mitra_Nematchoua_2024}. Scope 2 emissions represent the GHG associated with electricity demand by households and industry. Vulcan Scope 2 data are downscalings of US Balancing Authority estimates by de Chalendar and Benson \cite{deChalendar_Benson_2021}. They use economic input--output accounts to distribute electricity supply to household and industry sources of demand. Gurney et al. then downscale these estimates to census block groups using floor area and energy intensity factors \cite{Gurney_Dass_Kato_Mitra_Nematchoua_2024}. Census block group (CBG) estimates are validated by comparison with an EIA electricity consumption survey.

\begin{figure}[!ht]
    \centering
    \includegraphics[width=0.5\textwidth]{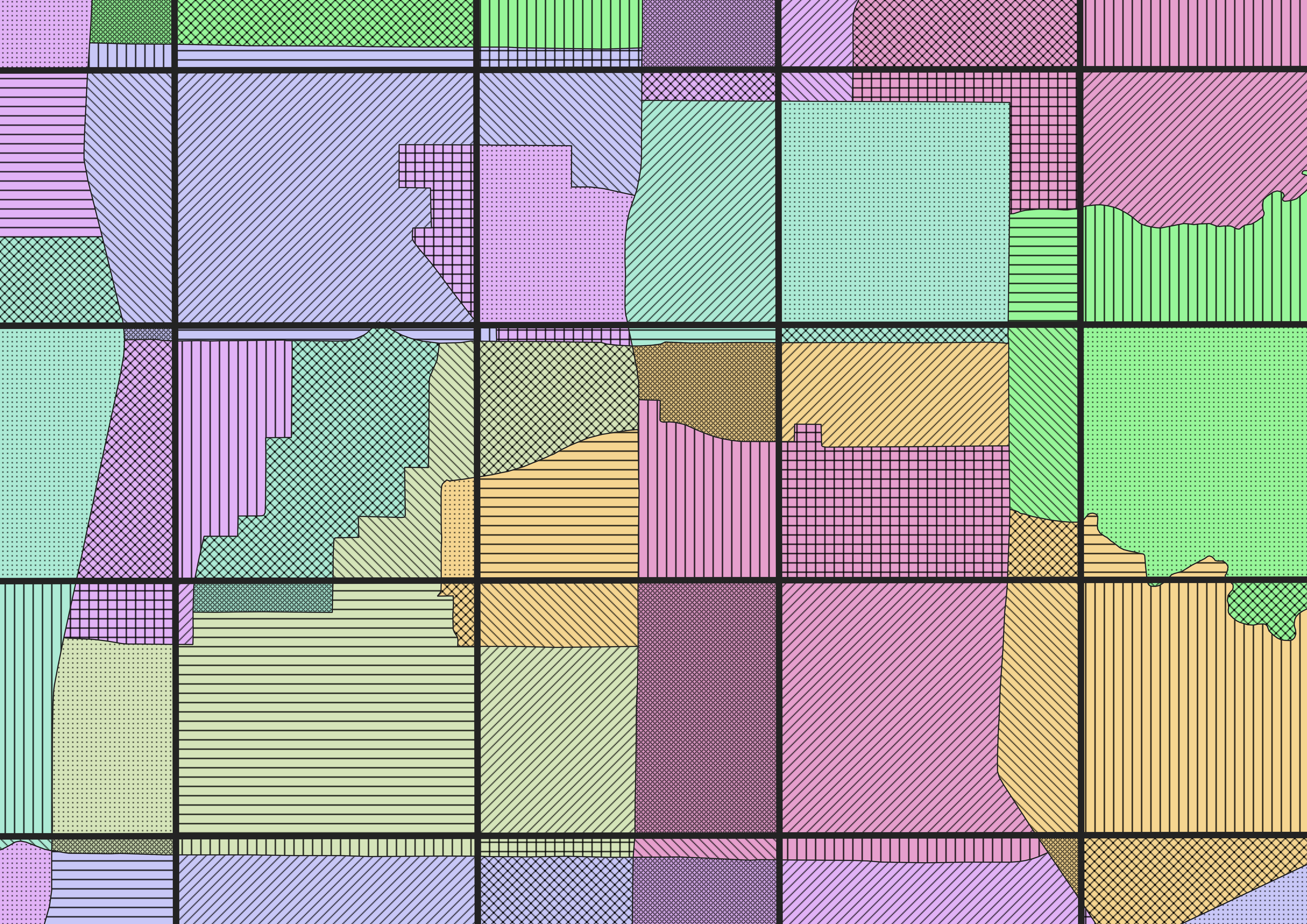}
    \caption{Gridcell-CBG spatial imputation}
    \label{fig:grid_cbg_imputation}
\end{figure}

A second core dataset to my analysis is the EPA Smart Location Database (EPA-SLD) \cite{Chapman_Fox_Bachman_Frank_Thomas_RourkReyes_2021}. This dataset summarizes CBG-level statistics related to demographics, employment, and built form variables for the United States, including measures of the 5Ds of land use. The use of publicly available datasets (such as Vulcan and EPA-SLD) facilitates open science and extension of the proposed work by other researchers. The spatial analysis unit is CBGs, such that Vulcan v4.0 gridcell data must be re-allocated among CBGs. This allocation is based on an areal interpolation and assumption of homogeneity within gridcells. For example, a CBG containing two gridcells is assigned the weighted average of their statistic values as a function of the proportion of each gridcell area contained within that CBG. Figure \ref{fig:grid_cbg_imputation} illustrates the process and demonstrates that gridcells are often coincident with CBG. I employed a similar approach in previous work on household consumption in the US to generate a uniform grid population density measure \cite{Hawkins_Habib_2021}. Areal interpolation is a common technique in spatial analysis fields \cite{Goodchild_Anselin_Deichmann_1993,Tsutsumi_Murakami_2014,Comber_Zeng_2019}. I chose to aggregate gridcells rather than CBGs because the aggregation affects GHG estimates only. In contrast, if I performed areal interpolation of CBG statistics to gridcells, I would have to account for both intensive (rate) and extensive (count) variables. 

The areal interpolation equation is as follows:

\begin{equation}
    CO2_{CBG} = \sum_i CO2_{i} \times \frac{Area_{i,CBG}}{Area_{i}}
\end{equation}

\noindent where $CO2_{CBG}$ is the interpolated CO2 for the CBG, $CO2_{i}$ is the CO2 for gridcell $i$ obtained from Vulcan v4.0, $Area_{i,CBG}$ is the area of gridcell $i$ in the CBG, and $Area_{i}$ is the total gridcell area (i.e., 1km). 

\subsection{Causal Identification Strategy}
\begin{figure}[!ht]
    \centering
    \includegraphics[width=0.9\textwidth]{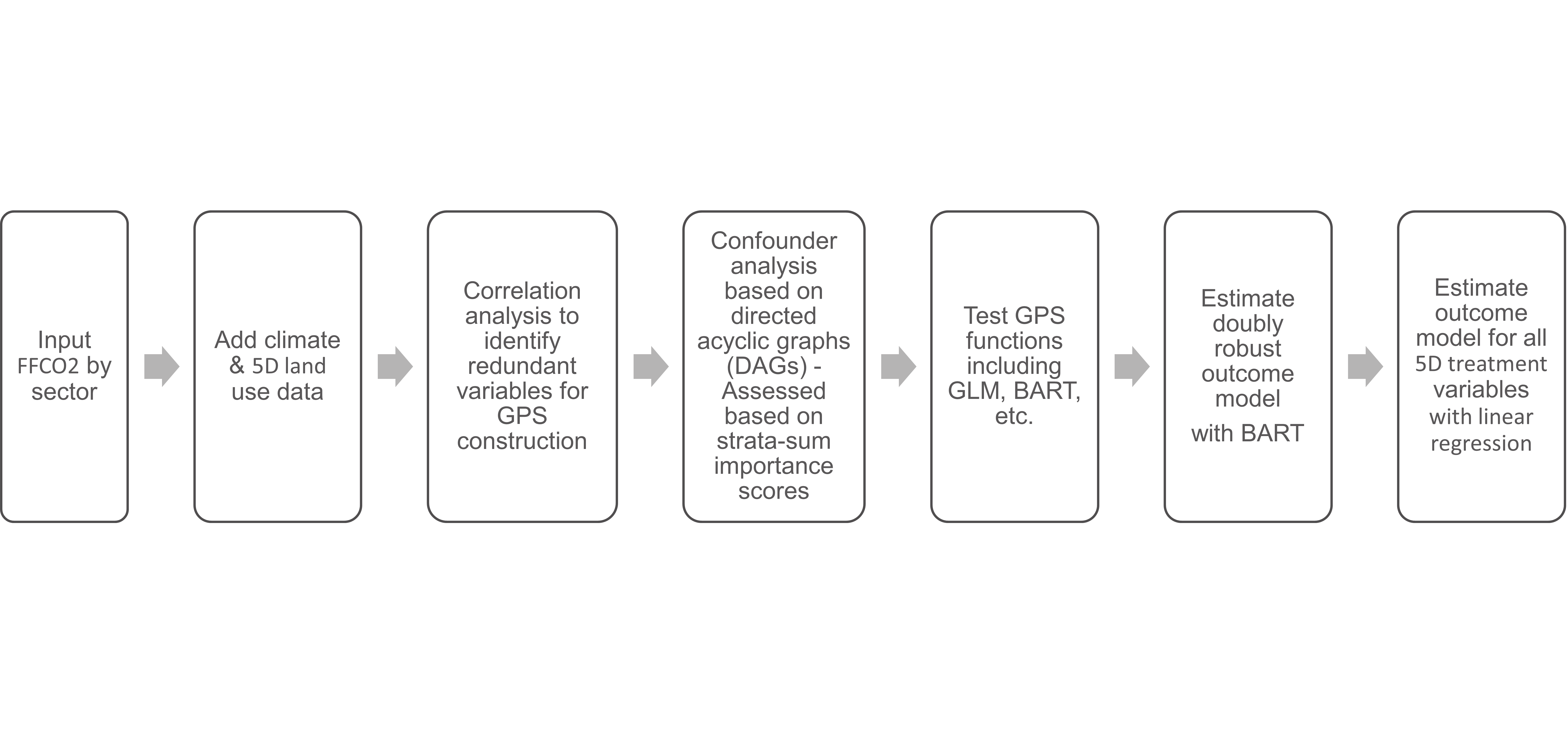}
    \caption{Overview of data and inference process}
    \label{fig:data_processing_flowchart}
\end{figure}

Using census block group (CBG)-level observations of emissions, I am interested in the causal effect of the 5D land use variables on FFCO2 emissions by sector. In this study, I focus on the emissions from transportation, residential electricity, and residential non-electricity energy sources. Among the causal identification strategies available for observation data, propensity score weighting is selected as the preferred option. This choice is motivated by several factors. The lack of a quasi-experimental interruption and panel observations of treatment variables limits the available options. The continuous treatments would be well suited to an instrumental variable (IV) approach but identifying multiple suitable variables is a barrier to implementing IV. Propensity score weighting (PSW) rather than matching (PSM) is suitable in this case because there are multiple continuous treatments. PSW lets us incorporate the generalised propensity scores (GPS) into either a doubly robust (DR) approach or regression model. The overall data and inference process is given in Figure \ref{fig:data_processing_flowchart}

My DR approach begins from a generalised propensity score, as proposed by Hirano and Imbens \cite{Hirano_Imbens_2005}. First, define $r(t,x)$ as the conditional density of the treatment $t$ given covariates $x$
\begin{equation}
    r(t,x) = f_{T|X}(t|x)
\end{equation}
and the generalised propensity score (GPS) as $R = r(T,X)$. The continuous GPS has similar properties to binary propensity scores; within strata of $r(t,X)$ the probability that $T=t$ does not depend on the value of $X$. The GPS then has the following property
\begin{equation}
    X \perp \mathbf{1}\{T=t\}|r(t,X)
\end{equation}
by the definition of the GPS. In combination with local weak unconfoundedness (i.e., unconfoundedness in the neighbourhood of a treatment level) conditional upon the treatment variables $X$, for every $t$ the following holds true
\begin{equation}
    f_T(t|r(t,X),Y(t)) = f_T(t|r(t,X))
\end{equation}
adapting the above notation to follow Graham \cite{Graham_2025}, the joint density of the observed data is then given by the following 
\begin{equation}
f_T(t) = f_{Y|T,X}(y|t,x)f_{T|X}(t|x)f_X(x)
\end{equation}

The DR estimator combines the outcome model $f_{Y|T,X}$ and GPS model $f_{T|X}$ to obtain a doubly robust average treatment effect. The estimator is doubly robust in that only one of the outcome and GPS models need be correctly specified.

The next challenge with continuous treatment analysis is how best to represent the average treatment effect (ATE). Let $t_{25}$ and $t_{75}$ denote the 25$^{th}$ and 75$^{th}$ percentile values of a continuous land use treatment. For unit $i$, the potential outcome under treatment level $t$ is $Y_i(t)$. The estimand used throughout the DR analysis is the interquartile average treatment effect

\begin{equation}
    ATE_{75,25} = E\left[Y(t_{75}) - Y(t_{25})\right]
\end{equation}
which is identified under weak unconfoundedness, $Y_i(t) \perp \mathbf{1}\{T_i=t\}|X_i$ for all treatment levels $t$ in the support of $T$, and continuous overlap, $0 < f_{T|X}(t|X_i) < \infty$ for $t \in \{t_{25},t_{75}\}$. This estimand compares two feasible points in the observed treatment distribution rather than a binary treated--untreated contrast

\begin{equation}
    \hat{\mu}(t_{25}) = \frac{1}{N} \sum_{i=1}^N \left[m(t_{25},X_i) + w_i(t_{25})\left\{Y_i - m(T_i,X_i)\right\}\right]
\end{equation}
\begin{equation}
    \hat{\mu}(t_{75}) = \frac{1}{N} \sum_{i=1}^N \left[m(t_{75},X_i) + w_i(t_{75})\left\{Y_i - m(T_i,X_i)\right\}\right]
\end{equation}
\begin{equation}
    \widehat{ATE}_{75,25} = \hat{\mu}(t_{75}) - \hat{\mu}(t_{25})
\end{equation}
where $\hat{\mu}(t_{25})$ and $\hat{\mu}(t_{75})$ are the doubly robust mean potential outcome estimators at the two treatment levels; $m(t,X_i)$ is the BART outcome model prediction; and $w_i(t)$ is the inverse-GPS weight constructed from $f_{T|X}(t|X_i)$, with the implementation using the estimated GPS at the evaluated treatment level. The use of 25$^{th}$ and 75$^{th}$ percentiles balances interpretability (interquartile range) with sufficient contrast to detect effects. The expected treatment effect is based on averaging over $N$ samples, but we can also visualise uncertainty in the ATE estimate via the posterior distribution given the use of BART.

\subsection{Confounder Variable Models}
The first step in the DR approach is to determine appropriate propensity score variables. Selection is based on a combination of available data, prior knowledge, and data-driven inference. In addition to land use variables available in the EPA Smart Location Database, I compile county-level climate data available from NOAA including precipitation, temperature minimum and maximum, and heating and cooling degree days. I also use 5-year (2018--2023) American Community Survey data to capture variations in sociodemographics. Residential energy and electricity demand are driven by heating and cooling needs and household income. Marginal FFCO2 emissions associated with electricity consumption are a function of the grid mix, which I capture through state-level eGrid emissions factors. Transportation emissions are correlated with mode choice through vehicle ownership, household size, and other demographic variables. They are correlated with VKT through income and other demographic variables. Variables for consideration in the propensity score are then filtered using correlation heatmap plots (see Supplementary Figs. 1, 2, and 3) to remove highly correlated variables. From these subsets of variables, I use a data-driven approach to select the propensity score variables for each treatment.

\begin{figure}[!ht]
    \centering
    \includegraphics{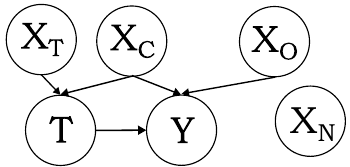}
    \caption{Causal graph for variable selection}
    \label{fig:causal_graph}
\end{figure}

Defining a causal graph (as in Figure \ref{fig:causal_graph}), covariates can be classified into four categories $X = \{X_C, X_T, X_O, X_N\}$ \cite{Tang_Kong_Pan_Wang_2023}. $X_C$ are confounders that affect both the treatment and outcome variables. $X_O$ and $X_T$ denote outcome and treatment predictors, respectively. $X_N$ denotes null variables that are irrelevant to the problem. $T$ and $Y$ are the treatment and outcome variable, respectively. Identifying the treatment effect relies on correctly identifying all confounding variables. Including treatment variables in the propensity score can cause bias and increased variance of the estimated treatment effect \cite{DeLuna_Waernbaum_Richardson_2011,Brookhart_Schneeweiss_Rothman_Glynn_Avorn_Stürmer_2006,Pearl_2011}, whereas null variables may function as colliders that should not be controlled in the propensity score. 

Following Li et al., I define a conditional independence test as defined in Algorithm \ref{alg:conditional_importance} \cite{Li_Pu_Cui_Lee_Guo_Ngoduy_2024}.

\begin{algorithm}[H]

\caption{Conditional Variable Importance Procedure for Continuous Treatments}
\label{alg:conditional_importance}
\begin{algorithmic}[1]

\Require Dataset with outcome $Y$, treatment $T$, covariates $X_1,\ldots,X_p$
\Require Number of treatment partitions $K$ (here $K=10$ deciles)
\Require Variable-importance threshold $\tau = 0.05$

\State Compute decile cutpoints of $T$ and assign each observation to a decile $D_i \in \{1,\ldots,K\}$

\For{$k = 1$ to $K$}
    \State Subset data to observations with $D_i = k$
    \State Fit a BART model: $Y_i = f_k(X_i) + \varepsilon_i$
    \For{each variable $X_j$}
        \State Compute conditional $R^2$ contribution $R^2_{jk}$ by
        pruning trees not containing $X_j$ and recomputing predictions
    \EndFor
\EndFor

\For{each variable $X_j$}
    \State Compute total conditional contribution:
    \[
        C_j = \sum_{k=1}^K R^2_{jk}
    \]
\EndFor

\State Define selected confounder set:
\[
    S = \{ X_j : C_j \ge \tau \}
\]

\State \Return Selected variable set $S$ and summary statistics for all $C_j$

\end{algorithmic}

\end{algorithm}

For their case, Li et al. use the Conditional Shapley Value Index (CSVI) as a test statistic defined for a binary treatment by \cite{Li_Pu_Cui_Lee_Guo_Ngoduy_2024} 
\begin{equation}
    CSVI(X,Y|T) = \omega Shap(X(1),\, Y(1)) + (1-\omega)Shap(X(0),\, Y(0)))
\end{equation}
where $\omega = P(T=1)$ is the probability of receiving the treatment and $Shap(X(t), Y(t))$ is the Shapley value of variable $X$ for continuous outcome $Y$. According to the definition of treatment predictors and null variables, $Shap(X(1),Y(1)) = Shap(X(0),Y(0))=0$ for these variables as the outcome $Y$ is unrelated to $X$ no matter the value of the treatment $T$. Li et al. use light Gradient Boosting Machines (GBM) to train the confounder variable model \cite{Li_Pu_Cui_Lee_Guo_Ngoduy_2024}. They set a threshold for CSVI to exclude variables with minimal conditional effect. I use Bayesian Additive Regression Trees (BART), described further below, for the confounder model. Unlike standard regression, BART captures nonlinear tipping points---such as the specific density thresholds at which emissions benefits plateau---without requiring manual specification of these complex interactions. A simple heuristic to quantify variable importance using BART is to count the number of times a variable is included in the regression trees. A plot of conditional prediction can be constructed by pruning trees from the BART model not containing the variable and calculating the corresponding R$^2$ value. However, the treatment in our case is continuous rather than binary. I approximate the CSVI approach by partitioning the treatment space into deciles, then estimating models conditional upon treatment variable range. This approach gives me marginal contributions to prediction accuracy for each variable and the ten deciles. I sum these marginal contributions based on intuition paralleling that of Li et al. \cite{Li_Pu_Cui_Lee_Guo_Ngoduy_2024}. Specifically, confounder variables should strongly describe the outcome conditional upon treatment decile (i.e., also describing treatment decile). I set a threshold of 0.05 representing variables describing at least 0.5\% of the outcome variation on average across deciles. This threshold was deemed conservative and tested during the propensity score balancing step. The considered confounder variables and their associated selection statistics are provided in Supplementary Table 1. Overall statistics are provided in Table \ref{tab:cvmodel_gof_summary}. Based on these statistics, a cut-off of 0.05 is a conservative variable inclusion criterion. Furthermore, Supplementary Table 9 shows the sensitivity of propensity score balancing models to marginal variable removal for electricity treatments. The reduced model for the density and diversity treatments increases the effective sample size (ESS) by 6\% and 0\%, respectively. Other summary statistics (coefficient of variation, MAD, and entropy) show variation but are within an acceptable range in all cases. Propensity score balance statistics show no correlation under either full variable or reduced models.

\begin{table}[h!]

\centering
\caption{Summary Statistics for Confounding Variable Model Conditional Goodness-of-Fit Results}
\label{tab:cvmodel_gof_summary}
\begin{tabular}{l r}
\toprule
\textbf{Statistic} & \textbf{Summation of $R^2$ Across Deciles} \\
\midrule
Minimum      & -0.021 \\
Maximum      & 8.247 \\
Mean     & 1.010 \\
Median   & 0.207 \\
10\%     & -0.004 \\
25\%     & 0.019 \\
75\%     & 0.966 \\
90\%     & 3.990 \\
\bottomrule
\end{tabular}

\end{table}

\subsection{Propensity Score Balancing}
I construct the propensity scores to ensure balance between treatment levels. Propensity score construction for continuous treatments is a challenging task. Many of the diagnostic tests developed for binary treatments are not applicable. Using the WeightIt package, I test parametric (CBPS) and nonparametric (NPCBPS) covariate balancing propensity score \cite{Fong_Hazlett_Imai_2018}, BART \cite{Hill_Weiss_Zhai_2011}, entropy balancing \cite{Vegetabile_Griffin_Coffman_Cefalu_Robbins_McCaffrey_2021, Tübbicke_2022}, generalised boosted models \cite{Zhu_Coffman_Ghosh_2015}, super learner \cite{Kreif_Grieve_Díaz_Harrison_2015}, and optimisation \cite{Greifer_2020} methods. Results are assessed for balance using propensity weighted Pearson correlation coefficients for confounder variables with the treatment. I also consider effective sample size, with a target minimum of 70 percent of the total observations. Balance diagnostics for several models are provided in Supplementary Information (SI) Section 2 (see Supplementary Tables 2-10). In all cases, standard mean differences (SMD) are less than 0.10 (< 0.05 in most cases) and many models achieve perfect balance across all confounder variables using either covariate balancing propensity scores (CBPS) or entropy balancing algorithms. Results for all variables (including additional algorithms and outcome variable transformations) are provided in code included with the Supplementary Information (SI).

In addition to confounder variables tested for all FFCO2 sources, I attempt to control for the biased transportation allocation by including two additional relevant confounders: annual FFCO2 generated by workers in a workplace CBG and commute vehicle miles travelled (VMT) per worker by CBG. These variables help account for differences in commuting VMT across CBG. 

\subsection{Outcome model}
A limitation of much causal research is its reliance on linear and logarithmic effect scaling \cite{Hill_2011}. While these are appealing assumptions in their simplicity, they may not adequately capture the non-linear relationship between the treatment variable of interest and the outcome. Machine learning approaches address linearity and parametric critiques but suffer from their own set of limitations arising from their lack of causal interpretation. Bayesian Additive Regression Trees (BART) is an approach that balances the flexibility of machine learning against the interpretability of classical causal econometric methods.

Causal inference in the absence of controlled or natural experimental data is a challenge. Many causal methods for observational data rely on modelling the treatment assignment, then the outcome variable conditional on its treatment assignment and confounding covariates. As demonstrated by Hill, BART provides a method to precisely model the outcome response surface \cite{Hill_2011}. 

The outcome model is specified for continuous land use treatments rather than binary treatment indicators. For each treatment $T$, I estimate the potential outcome response surface $Y_i(t)$ over observed treatment levels using the function $Y_i = f(T_i,X_i)+\epsilon_i$, where $X_i$ is the vector of confounding variables and $\epsilon_i$ are iid $N(0,\sigma^2)$ errors. Additive errors are assumed by BART, but the functional form of $f(T_i,X_i)$ is otherwise flexible. The causal contrast is therefore not $Y_i(1)-Y_i(0)$; it is the interquartile contrast $Y_i(t_{75})-Y_i(t_{25})$ defined in the causal identification section (4.1). Identification requires weak unconfoundedness of the potential outcomes at the evaluated continuous treatment levels and overlap in the GPS, so that the BART response surface is evaluated within the empirical support of the data. Because transportation emissions inherently involve spatial spillovers, classical SUTVA does not strictly hold in that sector; the spatial SLX+SEM models are used to explicitly represent this interference structure. This estimation becomes complex when $Y$ is not linearly related to $X$ and when the distribution of $X$ differs across treatment levels. Hill \cite{Hill_2011} argued that the simple implementation, precision, and robustness of BART estimates outweigh the benefits of alternative nonparametric approaches that are more strictly consistent under certain sets of regularity conditions.

The outcome model is defined by
\begin{equation}
\begin{split}
\ln(CO_2 \text{ per capita}) =& \text{ln(total CBSA population)} + \text{ln(CBSA population density)} \\&+ \text{ln(average CBG treatment level in CBSA)} + \text{propensity score} \\&+ \text{CBSA fixed effect}
\end{split}
\end{equation}
where CBSA is the census-based statistical area (CBSA) representing the metropolitan area. Consistent with the DR estimators defined in the causal identification section (4.2), the doubly robust average treatment effect is calculated as the difference between the estimated mean potential outcomes at the 75$^{th}$ and 25$^{th}$ percentile treatment levels. For each evaluated treatment level $t$, the doubly robust pseudo-outcome is
\begin{equation}
\hat{\mu}_{i}^{DR}(t) = m(t,X_i) + w_i(t)\left\{Y_i - m(T_i,X_i)\right\}
\end{equation}
where $m(t,X_i)$ is the BART prediction at treatment level $t$ and $w_i(t)$ is the inverse-GPS weight (as defined in section 4.2). The ATE is then the interquartile average difference $\frac{1}{N}\sum_i\left[\hat{\mu}_{i}^{DR}(t_{75})-\hat{\mu}_{i}^{DR}(t_{25})\right]$. Diagnostic tests for model convergence are provided in Supplementary Table 12.

\section*{Data Availability}
The datasets supporting the findings of this study are publicly available and can be accessed at https://zenodo.org/records/18895785. Vulcan data can be accessed at https://zenodo.org/records/15446748 (Scope 1) and https://zenodo.org/records/12123469 (Scope 2).

\section*{Code Availability}
Python (version 3.13) code and R (version 4.4) code were used to analyse and visualise the data. Code to reproduce the results is publicly available on GitHub (https://github.com/Hawkins-TECH-Lab/Carbon\_Burden\_Infrastructure\_US).

\section*{Acknowledgements}
This research did not receive any grant from funding agencies in the public, commercial, or not-for-profit sectors. The author is grateful for feedback received from three anonymous reviewers, presentation audiences at the NARSC annual meeting in San Diego, 7th Bridging Transport Researchers conference, 9th Conference on Econometric Models in Climate Change, and University of Calgary. He is also grateful to NPJ Sustainable Mobility and Transport for recognizing this research through the BTR7 Best Paper Award in Sustainable Transport.

\section*{Author Contributions}
J.H. has completed all sections of the manuscript.

\section*{Competing Interests}
The authors declare no competing financial or non-financial interests. 

\bibliography{references}

\newpage
\section{Propensity Score Confounding Variable Models}
\begin{figure}[!ht]
    \centering
    \includegraphics[width=0.75\textwidth]{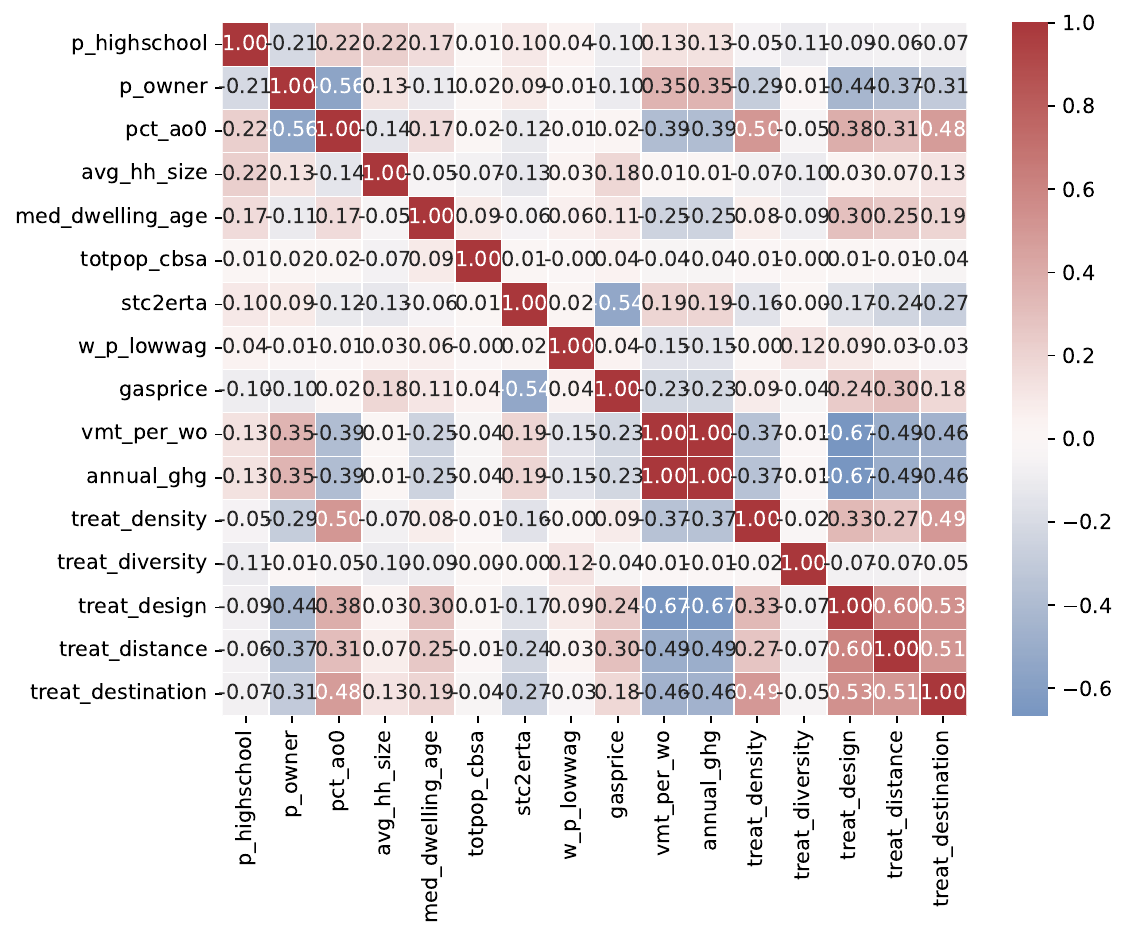}
    \caption{Transportation variable correlation heatmap}
    \label{fig:transport_heatmap}
\end{figure}

\begin{figure}[!ht]
    \centering
    \includegraphics[width=0.75\textwidth]{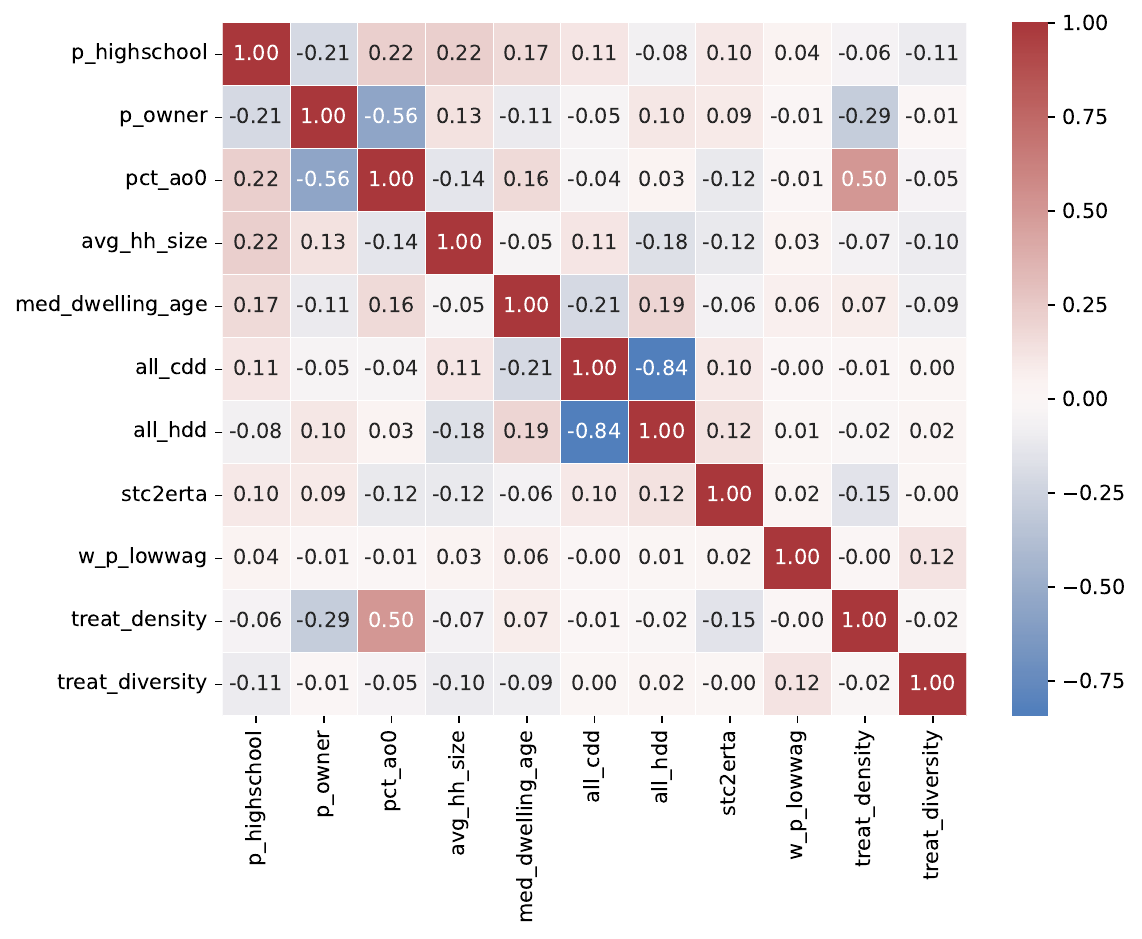}
    \caption{Residential electricity variable correlation heatmap}
    \label{fig:elec_heatmap}
\end{figure}

\begin{figure}[!ht]
    \centering
    \includegraphics[width=0.75\textwidth]{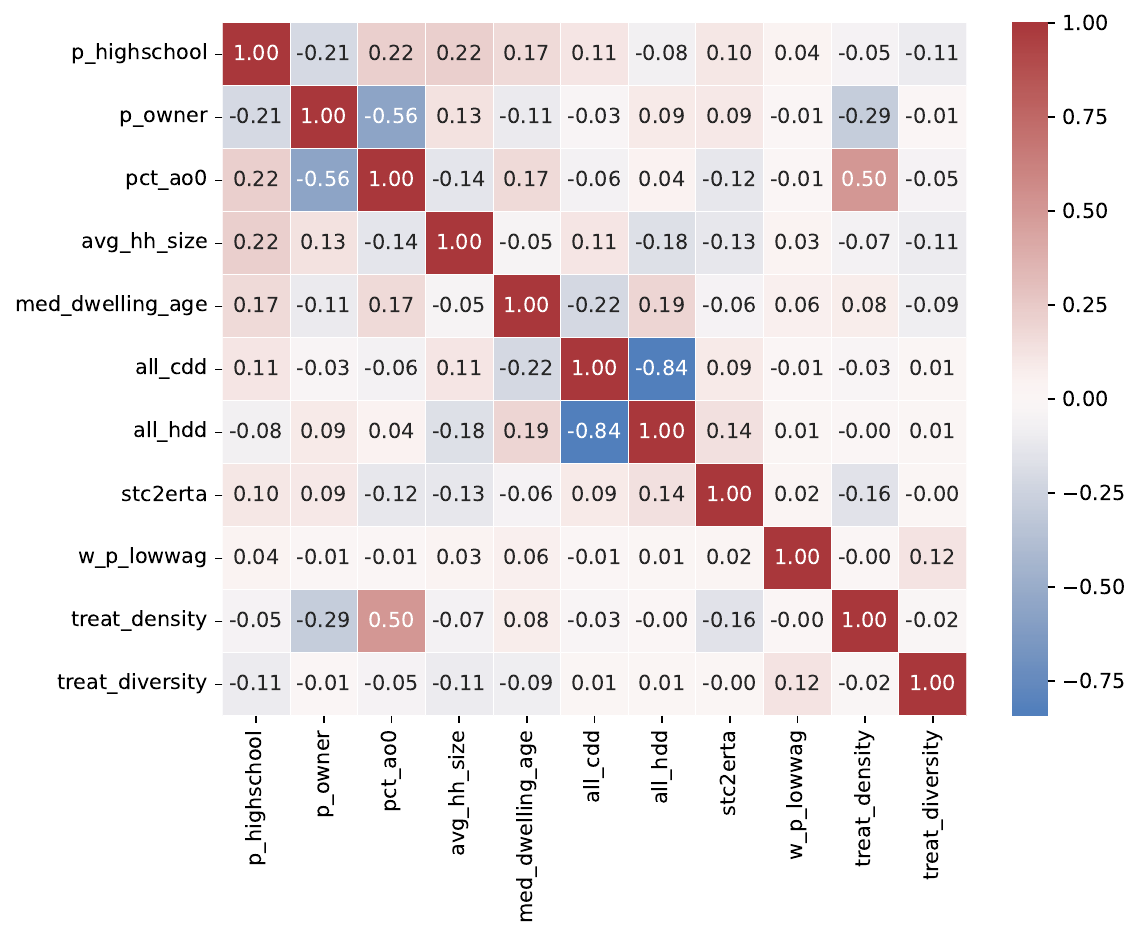}
    \caption{Residential energy variable correlation heatmap}
    \label{fig:residential_heatmap}
\end{figure}

I use BART models with 50 regression trees for all confounding-variable models. Convergence statistics were not assessed because goodness-of-fit statistics are used only for variable selection and not for model interpretation. Supplemental Figures~\ref{fig:transport_heatmap}--\ref{fig:residential_heatmap} show the correlation structure of the variables used in the transportation, residential electricity, and residential energy models.

\begin{longtable}{lccccc}
\caption{Confounding Variable Model Conditional Goodness-of-Fit Results}%
\label{tab:descr_stats}\\
\toprule
\textbf{Segment} & \textbf{Treatment} & \textbf{Feature} &
\multicolumn{3}{c}{\textbf{R\textsuperscript{2} Contribution by Treatment Decile}} \\
\cmidrule(l){4-6}
 & & & \textbf{Minimum} & \textbf{Median} & \textbf{Total} \\
\midrule
\endfirsthead

\caption[]{Confounding Variable Model Results (continued)}\\
\toprule
\textbf{Segment} & \textbf{Treatment} & \textbf{Feature} &
\multicolumn{3}{c}{\textbf{R\textsuperscript{2} Contribution by Treatment Decile}} \\
\cmidrule(l){4-6}
 & & & \textbf{Minimum} & \textbf{Median} & \textbf{Total} \\
\midrule
\endhead

\midrule
\multicolumn{6}{r}{\textit{Continued on next page}} \\
\endfoot

\bottomrule
\endlastfoot
Transportation          & Density     & med\_dwelling\_age & 0.14 & 0.47 & 4.33 \\
Transportation          & Density     & w\_p\_lowwag       & 0.06 & 0.08 & 0.97 \\
Transportation          & Density     & p\_highschool     & 0.01 & 0.05 & 0.58 \\
Transportation          & Density     & gasprice         & -0.01 & 0.02 & 0.55 \\
Transportation          & Density     & pct\_ao0          & -0.02 & 0.06 & 0.53 \\
Transportation          & Density     & stc2erta         & -0.01 & 0.03 & 0.33 \\
Transportation          & Density     & avg\_hh\_size      & -0.01 & 0.01 & 0.23 \\
Transportation          & Density     & totpop\_cbsa      & -0.01 & 0.00 & 0.00 \\
Transportation          & Density & VMT\_per\_wo      & -0.01 & 0.00 & 0.00 \\
Transportation          & Density & annual\_ghg      & -0.01 & 0.00 & 0.00 \\
Transportation          & Diversity   & med\_dwelling\_age & 0.06 & 0.17 & 2.09 \\
Transportation          & Diversity   & w\_p\_lowwag       & 0.03 & 0.09 & 1.03 \\
Transportation          & Diversity   & pct\_ao0          & -0.00 & 0.00 & 0.05 \\
Transportation          & Diversity   & gasprice         & -0.01 & -0.00 & -0.03 \\
Transportation          & Diversity   & p\_highschool     & -0.00 & -0.00 & -0.01 \\
Transportation          & Diversity   & stc2erta         & -0.01 & -0.00 & -0.02 \\
Transportation          & Diversity   & totpop\_cbsa      & -0.00 & 0.00 & 0.00 \\
Transportation          & Diversity   & avg\_hh\_size      & -0.01 & -0.00 & -0.01 \\
Transportation          & Diversity & VMT\_per\_wo      & -0.01 & 0.00 & 0.00 \\
Transportation          & Diversity & annual\_ghg      & -0.01 & 0.00 & 0.00 \\
Transportation          & Design & med\_dwelling\_age & 0.28 & 0.32 & 0.64 \\
Transportation          & Design & w\_p\_lowwag       & 0.03 & 0.07 & 0.14 \\
Transportation          & Design & stc2erta         & -0.00 & -0.00 & -0.01 \\
Transportation          & Design & gasprice         & -0.00 & -0.00 & -0.00 \\
Transportation          & Design & pct\_ao0          & 0.01 & 0.01 & 0.02 \\
Transportation          & Design & p\_highschool     & -0.00 & -0.00 & -0.00 \\
Transportation          & Design & avg\_hh\_size      & -0.00 & 0.00 & 0.00 \\
Transportation          & Design & totpop\_cbsa      & -0.00 & 0.00 & 0.00 \\
Transportation          & Design & VMT\_per\_wo      & -0.01 & 0.00 & 0.00 \\
Transportation          & Design & annual\_ghg      & -0.01 & 0.00 & 0.00 \\
Transportation          & Distance    & med\_dwelling\_age & 0.28 & 0.32 & 0.64 \\
Transportation          & Distance    & w\_p\_lowwag       & 0.03 & 0.07 & 0.14 \\
Transportation          & Distance    & pct\_ao0          & 0.01 & 0.01 & 0.02 \\
Transportation          & Distance    & gasprice         & -0.00 & -0.00 & -0.00 \\
Transportation          & Distance    & stc2erta         & -0.00 & -0.00 & -0.01 \\
Transportation          & Distance    & avg\_hh\_size      & -0.00 &  0.00 &  0.00 \\
Transportation          & Distance    & totpop\_cbsa      & -0.00 &  0.00 &  0.00 \\
Transportation          & Distance    & p\_highschool     & -0.00 & -0.00 & -0.00 \\
Transportation          & Distance & VMT\_per\_wo      & -0.01 & 0.00 & 0.00 \\
Transportation          & Distance & annual\_ghg      & -0.01 & 0.00 & 0.00 \\
Transportation          & Destination & med\_dwelling\_age & 0.06 & 0.34 & 2.81 \\
Transportation          & Destination & w\_p\_lowwag       & 0.00 & 0.04 & 0.75 \\
Transportation          & Destination & pct\_ao0          & -0.00 & 0.01 & 0.14 \\
Transportation          & Destination & stc2erta         & -0.00 & -0.00 & 0.03 \\
Transportation          & Destination & gasprice         & -0.01 & -0.00 & -0.03 \\
Transportation          & Destination & avg\_hh\_size      & -0.00 & 0.00 & 0.11 \\
Transportation          & Destination & p\_highschool     & -0.01 & -0.00 & -0.02 \\
Transportation          & Destination & totpop\_cbsa      & -0.01 & 0.00 & 0.00 \\
Transportation          & Destination & VMT\_per\_wo      & -0.01 & 0.00 & 0.00 \\
Transportation          & Destination & annual\_ghg      & -0.01 & 0.00 & 0.00 \\
Residential Energy      & Density     & all\_cdd                 & 0.77 & 0.83 & 8.25 \\
Residential Energy      & Density     & stc2erta                & 0.00 & 0.02 & 0.25 \\
Residential Energy      & Density     & p\_highschool            & 0.01 & 0.03 & 0.41 \\
Residential Energy      & Density     & pct\_ao0                 & -0.01 & 0.00 & 0.00 \\
Residential Energy      & Density     & avg\_hh\_size             & 0.00 & 0.00 & -0.01 \\
Residential Energy      & Density     & w\_p\_lowwag              & 0.00 & 0.00 & -0.01 \\
Residential Energy      & Density     & med\_dwelling\_age        & -0.01 & 0.01 & 0.10 \\
Residential Energy      & Diversity   & all\_cdd                 & 0.75 & 0.79 & 7.87 \\
Residential Energy      & Diversity   & p\_highschool            & 0.03 & 0.04 & 0.45 \\
Residential Energy      & Diversity   & w\_p\_lowwag              & 0.00 & 0.01 & 0.07 \\
Residential Energy      & Diversity   & stc2erta                & 0.01 & 0.01 & 0.15 \\
Residential Energy      & Diversity   & med\_dwelling\_age        & 0.00 & 0.02 & 0.27 \\
Residential Energy      & Diversity   & pct\_ao0                 & -0.01 & 0.00 & -0.02 \\
Residential Energy      & Diversity   & avg\_hh\_size             & -0.01 & 0.00 & 0.01 \\
Residential Electricity & Density     & med\_dwelling\_age & 0.01 & 0.48 & 3.81 \\
Residential Electricity & Density     & stc2erta         & 0.05 & 0.16 & 2.36 \\
Residential Electricity & Density     & all\_cdd          & 0.13 & 0.21 & 2.34 \\
Residential Electricity & Density     & p\_highschool     & 0.00 & 0.02 & 0.17 \\
Residential Electricity & Density     & pct\_ao0          & 0.00 & 0.00 & 0.04 \\
Residential Electricity & Density     & avg\_hh\_size      & -0.01 & 0.00 & 0.08 \\
Residential Electricity & Density     & w\_p\_lowwag       & -0.01 & 0.00 & -0.02 \\
Residential Electricity & Diversity   & med\_dwelling\_age & 0.30 & 0.50 & 4.60 \\
Residential Electricity & Diversity   & stc2erta         & 0.09 & 0.27 & 2.89 \\
Residential Electricity & Diversity   & all\_cdd          & 0.09 & 0.12 & 1.23 \\
Residential Electricity & Diversity   & pct\_ao0          & 0.01 & 0.01 & 0.12 \\
Residential Electricity & Diversity   & w\_p\_lowwag       & 0.00 & 0.00 & 0.01 \\
Residential Electricity & Diversity   & avg\_hh\_size      & -0.01 & 0.00 & 0.02 \\
Residential Electricity & Diversity   & p\_highschool     & 0.00 & 0.00 & 0.01 \\
\end{longtable}

\section{Propensity Score Balancing}\label{sec:psw_summary}
All algorithms use the default settings from the WeightIt package \citep{Greifer_2025}.

\subsection{Transportation}
\begin{table}[ht]
\small
\centering
\caption{Balance diagnostics across PSW algorithms for transportation density (dwelling units per square mile)}
\begin{tabular}{lccccccc}
\toprule
\textbf{\begin{tabular}[c]{@{}l@{}}Balance \\ Measures\end{tabular}} & \textbf{CBPS} & \textbf{BART} & \textbf{\begin{tabular}[c]{@{}c@{}}Entropy\\ Balancing\end{tabular}} & \textbf{\begin{tabular}[c]{@{}c@{}}Gradient \\ Boosting\\ Machine\end{tabular}} & \textbf{\begin{tabular}[c]{@{}c@{}}Super\\ Learner\end{tabular}} & \textbf{NPCBPS} & \textbf{Optimize$^*$} \\
\midrule
& \multicolumn{7}{c}{\textbf{Correlation}} \\
\midrule
Median dwelling age & 0 & -0.110 & 0.093 & 0.018 & 0.004 & 0.637 & 0.032 \\
\begin{tabular}[c]{@{}l@{}}Percent low wage \\ workers\end{tabular} & 0 & -0.812 & 0.037 & 0.017 & 0.008 & 0.431 & 0.018 \\
\begin{tabular}[c]{@{}l@{}}Gas price \\ (cents per gallon)\end{tabular} & 0 & -0.109 & 0.097 & 0.121 & -0.009 & -0.722 & 0.024 \\
\begin{tabular}[c]{@{}l@{}}Percent high school \\ or less\end{tabular} & 0 & 0.791 & -0.014 & -0.343 & 0.016 & 0.447 & 0.002 \\
\begin{tabular}[c]{@{}l@{}}Percent zero \\ vehicle households\end{tabular} & 0 & -2.035 & 0.182 & 0 & 0 & -2.026 & 0.040 \\
\begin{tabular}[c]{@{}l@{}}State grid CO2 \\ (lbs per MWh)\end{tabular} & 0 & 0.348 & -0.068 & -0.124 & 0 & 0.284 & -0.012 \\
\begin{tabular}[c]{@{}l@{}}Average household \\ size\end{tabular} & 0 & -0.230 & 0.002 & -0.239 & 0.030 & 0.361 & 0.008 \\
\midrule
\textbf{Sample size} & \multicolumn{7}{c}{Total = 216290} \\
\midrule
Effective sample size & 14772 & 9 & 186480 & 46173 & 1 & 1 & 162622 \\
\% Total & 7\% & 0\% & 86\% & 21\% & 0\% & 0\% & 75\% \\
\bottomrule
\end{tabular}

\begin{tablenotes}
\footnotesize
\item $^*$ denotes the selected PSW algorithm for the given treatment.
\end{tablenotes}
\end{table}

\begin{table}[ht]
\small
\centering
\caption{Balance diagnostics across PSW algorithms for land use diversity (5-digit NAICS code employment entropy)}
\begin{tabular}{lcccc}
\toprule
\textbf{\begin{tabular}[c]{@{}l@{}}Balance \\ Measures\end{tabular}} & \textbf{CBPS$^*$} & \textbf{BART} & \textbf{\begin{tabular}[c]{@{}c@{}}Entropy\\ Balancing\end{tabular}} & \textbf{\begin{tabular}[c]{@{}c@{}}Gradient \\ Boosting\\ Machine\end{tabular}} \\
\midrule
& \multicolumn{4}{c}{\textbf{Correlation}} \\
\midrule
Median dwelling age & 0 & -0.010 & 0 & -0.015 \\
\begin{tabular}[c]{@{}l@{}}Percent low wage \\ workers\end{tabular} & 0 & 0.005 & 0 & 0.010 \\
\begin{tabular}[c]{@{}l@{}}Percent zero \\ vehicle households\end{tabular} & 0 & -0.003 & 0 & -0.012 \\
\begin{tabular}[c]{@{}l@{}}Annual GHG emissions\end{tabular} & 0 & 0.002 & 0 & -0.002 \\
\begin{tabular}[c]{@{}l@{}}VMT per worker\end{tabular} & 0 & 0.002 & 0 & -0.002 \\
\midrule
\textbf{Sample size} & \multicolumn{4}{c}{Total = 216290} \\
\midrule
Effective sample size & 211546 & 140929 & 143031 & 154670 \\
\% Total & 98\% & 65\% & 66\% & 72\% \\
\bottomrule
\end{tabular}
\begin{tablenotes}
\footnotesize
\item $^*$ denotes the selected PSW algorithm for the given treatment.
\end{tablenotes}
\end{table}

\begin{table}[ht]
\small
\centering
\caption{Balance diagnostics across PSW algorithms for transportation design (facility miles per square mile)}
\begin{tabular}{lcccc}
\toprule
\textbf{\begin{tabular}[c]{@{}l@{}}Balance \\ Measures\end{tabular}} & \textbf{CBPS$^*$} & \textbf{BART} & \textbf{\begin{tabular}[c]{@{}c@{}}Entropy\\ Balancing\end{tabular}} & \textbf{\begin{tabular}[c]{@{}c@{}}Gradient \\ Boosting\\ Machine\end{tabular}} \\
\midrule
& \multicolumn{4}{c}{\textbf{Correlation}} \\
\midrule
Median dwelling age & 0 & -0.015 & 0 & -0.023 \\
Annual GHG emissions & 0 & 0.002 & 0 & 0.002 \\
VMT per worker & 0 & 0.002 & 0 & 0.002 \\
Percent low wage workers & 0 & 0 & 0 & 0.011 \\
Gas price (cents per gallon) & 0 & 0.013 & 0 & -0.002 \\
Percent high school or less & 0 & -0.016 & 0 & -0.020 \\
Percent zero vehicle households & 0 & -0.004 & 0 & -0.019 \\
State grid CO2 (lbs per MWh) & 0 & -0.013 & 0 & -0.016 \\
Average household size & 0 & 0.018 & 0 & -0.002 \\
\midrule
\textbf{Sample size} & \multicolumn{4}{c}{Total = 215788} \\
\midrule
Effective sample size & 206392 & 116659 & 142625 & 151281 \\
\% Total & 95\% & 54\% & 66\% & 70\% \\
\bottomrule
\end{tabular}

\begin{tablenotes}
\footnotesize
\item $^*$ denotes the selected PSW algorithm for the given treatment.
\end{tablenotes}
\end{table}

\begin{table}[ht]
\small
\centering
\caption{Balance diagnostics across PSW algorithms for distance to transit station (metres)}
\begin{tabular}{lcccc}
\toprule
\textbf{\begin{tabular}[c]{@{}l@{}}Balance \\ Measures\end{tabular}} & \textbf{CBPS} & \textbf{BART} & \textbf{\begin{tabular}[c]{@{}c@{}}Entropy$^*$\\ Balancing\end{tabular}} & \textbf{\begin{tabular}[c]{@{}c@{}}Gradient \\ Boosting\\ Machine\end{tabular}} \\
\midrule
& \multicolumn{4}{c}{\textbf{Correlation}} \\
\midrule
Median dwelling age & 0 & -0.031 & 0 & 0.032 \\
Percent low wage workers & 0 & -0.002 & 0 & 0.016 \\
Gas price (cents per gallon) & 0 & -0.050 & 0 & 0.020 \\
Percent zero vehicle households & 0 & 0.003 & 0 & 0.052 \\
\midrule
\textbf{Sample size} & \multicolumn{4}{c}{Total = 216290} \\
\midrule
Effective sample size & 85160 & 116337 & 141438 & 151281 \\
\% Total & 39\% & 54\% & 65\% & 77\% \\
\bottomrule
\end{tabular}

\begin{tablenotes}
\footnotesize
\item $^*$ denotes the selected PSW algorithm for the given treatment.
\end{tablenotes}
\end{table}

\begin{table}[t]
\small
\centering
\caption{Balance diagnostics across PSW algorithms for destination accessibility (jobs within 45 minutes auto travel time, time-decay
(network travel time) weighted)}
\begin{tabular}{lcccc}
\toprule
\textbf{\begin{tabular}[c]{@{}l@{}}Balance \\ Measures\end{tabular}} & \textbf{CBPS$^*$} & \textbf{BART} & \textbf{\begin{tabular}[c]{@{}c@{}}Entropy\\ Balancing\end{tabular}} & \textbf{\begin{tabular}[c]{@{}c@{}}Gradient \\ Boosting\\ Machine\end{tabular}} \\
\midrule
& \multicolumn{4}{c}{\textbf{Correlation}} \\
\midrule
Median dwelling age & 0 & -0.099 & 0.000 & 0.115 \\
Percent low wage workers & 0 & 0.031 & 0.000 & -0.031 \\
Gas price (cents per gallon) & 0 & 0.057 & -0.003 & 0.143 \\
Percent high school or less & 0 & -0.007 & -0.008 & -0.099 \\
Percent zero vehicle households & 0 & -0.503 & 0.001 & 0.241 \\
State grid CO2 (lbs per MWh) & 0 & 0.039 & -0.003 & -0.182 \\
\midrule
\textbf{Sample size} & \multicolumn{4}{c}{Total = 215788} \\
\midrule
Effective sample size & 152388 & 47 & 92054 & 196296 \\
\% Total & 70\% & 0\% & 43\% & 91\% \\
\bottomrule
\end{tabular}

\begin{tablenotes}
\footnotesize
\item $^*$ denotes the selected PSW algorithm for the given treatment.
\end{tablenotes}
\end{table}

\FloatBarrier

\subsection{Residential Electricity}
\begin{table}[ht]
\small
\centering
\caption{Balance diagnostics across PSW algorithms for population density (dwelling units per square mile)}
\begin{tabular}{lcccccc}
\toprule
\textbf{\begin{tabular}[c]{@{}l@{}}Balance \\ Measures\end{tabular}} & \textbf{CBPS} & \textbf{BART} & \textbf{\begin{tabular}[c]{@{}c@{}}Entropy$^*$\\ Balancing\end{tabular}} & \textbf{\begin{tabular}[c]{@{}c@{}}Gradient \\ Boosting\\ Machine\end{tabular}} & \textbf{NPCBPS} & \textbf{Optimize} \\
\midrule
& \multicolumn{6}{c}{\textbf{Correlation}} \\
\midrule
Median dwelling age & 0 & -0.095 & 0.049 & 0.013 & 0.115 & 0.198 \\
State grid CO2 (lbs per MWh) & 0 & 0.214 & -0.053 & -0.170 & -0.113 & -0.234 \\
Annual cooling degree days & 0 & 0.015 & 0.005 & 0.262 & 0.002 & -0.035 \\
Percent zero vehicle households & 0 & -0.746 & 0.077 & 0 & 0.112 & 0.568 \\
\midrule
\textbf{Sample size} & \multicolumn{6}{c}{Total = 217182} \\
\midrule
Effective sample size & 37767 & 8 & 171379 & 50024 & 1194 & 217182 \\
\% Total & 17\% & 0\% & 71\% & 23\% & 1\% & 100\% \\
\bottomrule
\end{tabular}

\begin{tablenotes}
\footnotesize
\item $^*$ denotes the selected PSW algorithm for the given treatment.
\end{tablenotes}
\end{table}
\FloatBarrier

\begin{table}[ht]
\small
\centering
\caption{Balance diagnostics across PSW algorithms for land use diversity (5-digit NAICS code employment entropy)}
\begin{tabular}{lcccc}
\toprule
\textbf{\begin{tabular}[c]{@{}l@{}}Balance \\ Measures\end{tabular}} & \textbf{CBPS$^*$} & \textbf{BART} & \textbf{\begin{tabular}[c]{@{}c@{}}Entropy\\ Balancing\end{tabular}} & \textbf{\begin{tabular}[c]{@{}c@{}}Gradient \\ Boosting\\ Machine\end{tabular}} \\
\midrule
& \multicolumn{4}{c}{\textbf{Correlation}} \\
\midrule
Median dwelling age & 0 & -0.001 & 0 & -0.005 \\
State grid CO2 (lbs per MWh) & 0 & 0.001 & 0 & 0.001 \\
Annual cooling degree days & 0 & -0.003 & 0 & -0.004 \\
Percent zero vehicle households & 0 & 0.002 & 0 & -0.001 \\
\midrule
\textbf{Sample size} & \multicolumn{4}{c}{Total = 217182} \\
Effective sample size & 215110 & 206419 & 213960 & 211350 \\
\% Total & 99\% & 95\% & 99\% & 97\% \\
\bottomrule
\end{tabular}

\begin{tablenotes}
\footnotesize
\item $^*$ denotes the selected PSW algorithm for the given treatment.
\end{tablenotes}
\end{table}

\begin{table}[h!]
\centering
\captionsetup{width=\textwidth}
\caption{Comparison of Reduced Models and Full Variable Sets for Electricity Density and Electricity Diversity}
\label{tab:electricity_density_diversity}
\begin{tabular}{lrrr}
\toprule
\textbf{Metric} & \textbf{Reduced Model} & \textbf{Full Variable Set} & \textbf{Difference} \\
\midrule
\multicolumn{4}{l}{\textbf{Electricity Density} (removing \textit{avg\_hh\_size}, conditional $R^2 = 0.080$)} \\
Coef.\ of Variation (weights) & 0.478 & 0.546 & $-12\%$ \\
MAD (weights)                 & 0.235 & 0.257 & $-9\%$ \\
Entropy                       & 0.077 & 0.094 & $-18\%$ \\
ESS (weighted)                & 176{,}367.70 & 166{,}923.90 & $6\%$ \\
\midrule
\multicolumn{4}{l}{\textbf{Electricity Diversity} (removing \textit{pct\_ao0}, conditional $R^2 = 0.119$)} \\
Coef.\ of Variation           & 0.093 & 0.101 & $-8\%$ \\
MAD                           & 0.064 & 0.067 & $-4\%$ \\
Entropy                       & 0.004 & 0.005 & $-20\%$ \\
ESS (weighted)                & 214{,}756.9 & 214{,}451.7 & $0\%$ \\
\bottomrule
\end{tabular}
\end{table}

\FloatBarrier
\newpage

\subsection{Residential Energy}
\begin{table}[ht]
\small
\centering
\caption{Balance diagnostics across PSW algorithms for population density (dwelling units per square mile)}
\begin{tabular}{lcccc}
\toprule
\textbf{\begin{tabular}[c]{@{}l@{}}Balance \\ Measures\end{tabular}} & \textbf{CBPS$^*$} & \textbf{BART} & \textbf{\begin{tabular}[c]{@{}c@{}}Entropy\\ Balancing\end{tabular}} & \textbf{\begin{tabular}[c]{@{}c@{}}Gradient \\ Boosting\\ Machine\end{tabular}} \\
\midrule
& \multicolumn{4}{c}{\textbf{Correlation}} \\
\midrule
Percent high school or less & 0 & -0.187 & -0.002 & -0.093 \\
Median dwelling age & 0 & -0.009 & 0.027 & 0.000 \\
State grid CO2 (lbs per MWh) & 0 & 0.175 & -0.050 & -0.050 \\
Annual cooling degree days & 0 & 0.033 & 0.005 & 0.028 \\
\midrule
\textbf{Sample size} & \multicolumn{4}{c}{Total = 216230} \\
\midrule
Effective sample size & 178071 & 47 & 194601 & 134302 \\
\% Total & 7\% & 0\% & 39\% & 91\% \\
\bottomrule
\end{tabular}

\begin{tablenotes}
\footnotesize
\item $^*$ denotes the selected PSW algorithm for the given treatment.
\end{tablenotes}
\end{table}

\begin{table}[ht]
\small
\centering
\caption{Balance diagnostics across PSW algorithms for land use diversity (5-digit NAICS code employment entropy)}
\begin{tabular}{lcccc}
\toprule
\textbf{\begin{tabular}[c]{@{}l@{}}Balance \\ Measures\end{tabular}} & \textbf{CBPS$^*$} & \textbf{BART} & \textbf{\begin{tabular}[c]{@{}c@{}}Entropy\\ Balancing\end{tabular}} & \textbf{\begin{tabular}[c]{@{}c@{}}Gradient \\ Boosting\\ Machine\end{tabular}} \\
\midrule
& \multicolumn{4}{c}{\textbf{Correlation}} \\
\midrule
State grid CO2 (lbs per MWh) & 0 & -0.007 & 0 & -0.008 \\
Annual cooling degree days & 0 & -0.003 & 0.003 & -0.001 \\
Percent high school or less & 0 & -0.013 & -0.008 & -0.019 \\
Percent low wage workers & 0 & 0.004 & -0.011 & 0.014 \\
\midrule
\textbf{Sample size} & \multicolumn{4}{c}{Total = 216230} \\
\midrule
Effective sample size & 209646 & 130342 & 185182 & 152566 \\
\% Total & 98\% & 60\% & 86\% & 86+\% \\
\bottomrule
\end{tabular}

\begin{tablenotes}
\footnotesize
\item $^*$ denotes the selected PSW algorithm for the given treatment.
\end{tablenotes}
\end{table}

\FloatBarrier

\section{Doubly Robust BART Model}
I use BART models with 25 regression trees for all doubly robust outcome models. Standard Bayesian model statistics are not easily interpretable for BART models. I use percentiles of the $\hat{r}$ statistic to summarize the distribution of convergence statistics for regression-tree parameters. The 90\% percentiles of the $\hat{r}$ statistic are less than 1.35 for all models, with most statistics less than 1.10, suggesting reasonably good model convergence.

\begin{table}[ht]
\caption{Summary of BART Model Convergence Diagnostics ($\hat{r}$ percentiles)}
\label{tab:bart_convergence}
\begin{tabular}{@{}lccc@{}}
\toprule
                        & \multicolumn{3}{c}{\textbf{Percentiles}} \\ \midrule
                        & 10\%         & 50\%        & 90\%        \\
\textbf{Transportation} &              &             &             \\
Density                 & 1.21        & 1.23        & 1.25       \\
Diversity               & 1.03        & 1.05       & 1.06        \\
Design                  & 1.01        & 1.01       & 1.02       \\
Distance                & 1.02         & 1.02       & 1.03       \\
Destination             & 1.06        & 1.07       & 1.08       \\
\textbf{Electricity}    &              &             &             \\
Density                 & 1.07        & 1.08       & 1.09       \\
Diversity               & 1.03        & 1.04       & 1.05       \\
\textbf{Residential}    &              &             &             \\
Density                 & 1.08         & 1.09       & 1.11       \\
Diversity               & 1.04            & 1.05          & 1.05           \\ \bottomrule
\end{tabular}
\end{table}

\FloatBarrier

\section{Spatial Robustness Checks}

\begin{table}[htbp]
\centering
\caption{Joint treatment effects of land use variables on transportation FFCO2 emissions}
\label{tab:gmswls_three_models_tran}
\small
\begin{tabular}{@{}lcccccc@{}}
\toprule
 & \multicolumn{2}{c}{Spatial Error Weights}
 & \multicolumn{2}{c}{KNN Weights (k=10)}
 & \multicolumn{2}{c}{Inverse Exponential Weights} \\
Variable 
& Coef. & $p$
& Coef. & $p$
& Coef. & $p$ \\
\midrule
Constant 
& 1.252 & 0.00
& 1.303 & 0.00
& 1.517 & 0.00 \\

Total population (CBSA)
& -0.029 & 0.61
& -0.024 & 0.70
& -0.045 & 0.42 \\

Average population density (CBSA)
& 0.288 & 0.41
& 0.260 & 0.49
& 0.172 & 0.61 \\

Average land use diversity (CBSA)
& 0.004 & 0.69
& 0.011 & 0.27
& 0.018 & 0.06 \\

Average roadway design (CBSA)
& -0.012 & 0.41
& -0.009 & 0.54
& -0.017 & 0.24 \\

Average distance to transit (CBSA)
& 0.002 & 0.79
& 0.004 & 0.62
& 0.001 & 0.91 \\

Average destination access (CBSA)
& 0.007 & 0.90
& 0.020 & 0.72
& 0.069 & 0.18 \\

Population density PS
& -0.104 & 0.00
& -0.125 & 0.00
& -0.113 & 0.00 \\

Land-use diversity PS
& -0.005 & 0.11
& -0.005 & 0.11
& 0.004 & 0.22 \\

Roadway design PS
& 0.006 & 0.01
& 0.002 & 0.53
& 0.001 & 0.79 \\

Distance to transit PS
& 0.022 & 0.00
& 0.027 & 0.00
& 0.025 & 0.00 \\

Destination accessibility PS
& 0.029 & 0.00
& 0.030 & 0.00
& 0.037 & 0.00 \\

Population density (CBG)
& -0.194 & 0.00
& -0.213 & 0.00
& -0.191 & 0.00 \\

Land-use diversity (CBG)
& 0.081 & 0.00
& 0.105 & 0.00
& 0.114 & 0.00 \\

Roadway design (CBG)
& -0.450 & 0.00
& -0.517 & 0.00
& -0.460 & 0.00 \\

Distance to transit (CBG)
& 0.033 & 0.00
& 0.041 & 0.00
& 0.040 & 0.00 \\

Destination accessibility (CBG)
& 0.347 & 0.00
& 0.408 & 0.00
& 0.401 & 0.00 \\

Density × PS (CBG)
& 0.039 & 0.00
& 0.046 & 0.00
& 0.033 & 0.00 \\

Diversity × PS (CBG)
& 0.019 & 0.00
& 0.020 & 0.00
& 0.027 & 0.00 \\

Design × PS (CBG)
& 0.009 & 0.00
& 0.012 & 0.00
& 0.008 & 0.00 \\

Distance × PS (CBG)
& 0.006 & 0.00
& 0.006 & 0.01
& 0.001 & 0.77 \\

Destination × PS (CBG)
& 0.004 & 0.03
& 0.008 & 0.00
& 0.013 & 0.00 \\

\midrule

$\lambda$
& \multicolumn{2}{c}{0.656}
& \multicolumn{2}{c}{0.701}
& \multicolumn{2}{c}{0.595} \\

Pseudo $R^2$
& \multicolumn{2}{c}{0.260}
& \multicolumn{2}{c}{0.257}
& \multicolumn{2}{c}{0.260} \\

Residual Moran I 
& 0.380 & 0.00 
& 0.264 & 0.00 
& 0.170 & 0.00 \\

\bottomrule
\end{tabular}
\footnotesize{\textit{Notes:} CBSA fixed effects omitted for clarity.}
\end{table}

\begin{table}[htbp]
\centering
\caption{Joint treatment effects of land use variables on residential electricity FFCO2 emissions}
\label{tab:gmswls_three_models_elec}
\small
\begin{tabular}{@{}lcccccc@{}}
\toprule
 & \multicolumn{2}{c}{Spatial Error Weights}
 & \multicolumn{2}{c}{KNN Weights (k=10)}
 & \multicolumn{2}{c}{Inverse Exponential Weights} \\
Variable 
& Coef. & $p$
& Coef. & $p$
& Coef. & $p$ \\
\midrule
Constant 
& -0.227 & 0.00
& -0.153 & 0.00
& -0.144 & 0.00 \\

Total population (CBSA)
& -0.020 & 0.14
& -0.032 & 0.04
& -0.030 & 0.04 \\

Average population density (CBSA)
& -0.011 & 0.92
& 0.181 & 0.16
& 0.100 & 0.42 \\

Average land-use diversity (CBSA)
& -0.012 & 0.01
& -0.007 & 0.13
& -0.016 & 0.00 \\

Population density PS
& 0.012 & 0.00
& 0.007 & 0.00
& 0.012 & 0.00 \\

Land-use diversity PS
& -0.009 & 0.00
& -0.011 & 0.00
& -0.012 & 0.00 \\

Population density (CBG)
& -0.061 & 0.00
& -0.062 & 0.00
& -0.056 & 0.00 \\

Land-use diversity (CBG)
& 0.010 & 0.00
& 0.008 & 0.00
& 0.013 & 0.00 \\

Population density × PS (CBG)
& 0.000 & 0.43
& 0.001 & 0.19
& -0.000 & 0.69 \\

Land-use diversity × PS (CBG)
& -0.012 & 0.00
& -0.014 & 0.00
& -0.018 & 0.00 \\

\midrule
Weights Matrix 
& \multicolumn{2}{c}{GM-SWLS (Spec 1)} 
& \multicolumn{2}{c}{GM-SWLS (Spec 2)}
& \multicolumn{2}{c}{GM-SWLS (Spec 3)} \\

$\lambda$
& \multicolumn{2}{c}{0.619}
& \multicolumn{2}{c}{0.737}
& \multicolumn{2}{c}{0.757} \\

Pseudo $R^2$
& \multicolumn{2}{c}{0.643}
& \multicolumn{2}{c}{0.639}
& \multicolumn{2}{c}{0.631} \\

Residual Moran I 
& 0.451 & 0.00 
& 0.407 & 0.00 
& 0.363 & 0.00 \\

\bottomrule
\end{tabular}
\footnotesize{\textit{Notes:} CBSA fixed effects omitted for clarity.}
\end{table}

\begin{table}[htbp]
\centering
\caption{Joint treatment effects of land use variables on residential energy FFCO2 emissions}
\label{tab:gmswls_three_models_res_energy}
\small
\begin{tabular}{@{}lcccccc@{}}
\toprule
 & \multicolumn{2}{c}{Spatial Error Weights}
 & \multicolumn{2}{c}{KNN Weights (k=10)}
 & \multicolumn{2}{c}{Inverse Exponential Weights} \\
Variable 
& Coef. & $p$
& Coef. & $p$
& Coef. & $p$ \\
\midrule
Constant 
& -1.086 & 0.00
& -1.108 & 0.00
& -1.114 & 0.00 \\

Total population (CBSA)
& 0.021 & 0.28
& 0.015 & 0.47
& 0.027 & 0.19 \\

Average population density (CBSA)
& 0.056 & 0.73
& 0.145 & 0.42
& 0.019 & 0.91 \\

Average land-use diversity (CBSA)
& -0.007 & 0.22
& -0.007 & 0.27
& -0.010 & 0.15 \\

Population density PS
& 0.008 & 0.00
& 0.004 & 0.05
& 0.007 & 0.00 \\

Land-use diversity PS
& 0.015 & 0.00
& 0.016 & 0.00
& 0.013 & 0.00 \\

Population density (CBG)
& -0.157 & 0.00
& -0.169 & 0.00
& -0.149 & 0.00 \\

Land-use diversity (CBG)
& 0.001 & 0.60
& 0.002 & 0.13
& 0.002 & 0.26 \\

Population density × PS (CBG)
& -0.023 & 0.00
& -0.018 & 0.00
& -0.022 & 0.00 \\

Land-use diversity × PS (CBG)
& 0.011 & 0.00
& 0.012 & 0.00
& 0.009 & 0.00 \\

\midrule
Weights Matrix 
& \multicolumn{2}{c}{GM-SWLS (Model 1)} 
& \multicolumn{2}{c}{GM-SWLS (Model 2)}
& \multicolumn{2}{c}{GM-SWLS (Model 3)} \\

$\lambda$
& \multicolumn{2}{c}{0.483}
& \multicolumn{2}{c}{0.598}
& \multicolumn{2}{c}{0.570} \\

Pseudo $R^2$
& \multicolumn{2}{c}{0.637}
& \multicolumn{2}{c}{0.635}
& \multicolumn{2}{c}{0.636} \\

Residual Moran I 
& 0.226 & 0.00 
& 0.264 & 0.00 
& 0.170 & 0.00 \\

\bottomrule
\end{tabular}
\footnotesize{\textit{Notes:} CBSA fixed effects omitted for clarity.}
\end{table}

\FloatBarrier

\section{Vulcan Greenhouse Gas Emissions Estimates}
The Vulcan dataset comprises 1-km gridcell estimates of fossil-fuel CO$_2$ (FFCO2) for the United States at one-hour intervals. Vulcan 1.0 was developed for 10-km gridcell estimates for the year 2002. The current version is Vulcan v4.0, which includes annual estimates for the years 2010--2021. The dataset uses a bottom-up approach based on point-source reporting to the Environmental Protection Agency (EPA), line and polygon features, and numerous statistical reports of energy consumption in the transportation, residential, commercial, industrial, and other sectors. Vulcan v4.0 disaggregates emissions for the following sectors:
\begin{itemize}
    \item Residential
    \item Commercial
    \item Industrial
    \item Electricity production
    \item Onroad
    \item Nonroad
    \item Airport
    \item Rail
    \item Commercial marine vessels
    \item Cement
\end{itemize}

Vulcan estimates were compared with US estimates from the ODIAC global data product \citep{Oda_Maksyutov_Andres_2018}. Gurney et al. found large differences in both total emissions (100.1 TgC for 2011) and spatial distribution (spatial correlation of 0.38) \citep{Gurney_Liang_Patarasuk_Song_Huang_Roest_2020}. However, a comparison with $^{14}$CO$_2$ measurements and an atmospheric inversion approach by Basu et al. \citep{basu_estimating_2020} found that estimates differed by only 1.4\%.

\subsection{Scope 2 Emissions Estimates}
Scope 2 greenhouse gas (GHG) emissions estimates, in units of CO$_2$-equivalent emissions and electricity consumption, are taken from Gurney et al. for census block groups (BGs). Their estimates are based on estimates from Chalendar et al. for US Balancing Authority (BA) areas. The mixture of power plants providing electricity for each BA is based on multiregional input-output (MRIO) tables. Emission factors used in the calculations vary by season, location, and time of day. Gurney et al. downscale these data from the BA-scale to BGs using energy and infrastructure proxy variables: total building floor area categorized by building type and electricity energy use intensity (EUI) for each building type present in a BG. Estimates are validated using self-reported utility electricity consumption data archived by the Energy Information Administration (EIA). A total of 185 utilities met the isolation criteria defined by Gurney et al. Total electricity consumption had a median relative difference of 13.4\%, measured as the difference between reported utility-level consumption and aggregations of BG Vulcan data to utility service areas. The residential-sector median relative difference was -10.4\%. Both the Vulcan and EIA data likely represent biased measures relative to true electricity consumption. However, these comparisons validate that the Vulcan estimates are aligned with available aggregate statistics.

\end{document}